\documentclass[]{aa}  
\usepackage{graphicx}
\usepackage{float}
\usepackage{xcolor}
\usepackage{natbib}
\DeclareUnicodeCharacter{00A0}{~}

\def \MS{M$\mathrm{_\odot}$}

\def \kms{km\,s$^{-1}$}

\def \1s{$1\,\sigma$}

\def \t0{T$_0$}

\defcitealias{Millholland2018}{M18}

\def\rebound{\texttt{Rebound}}
\def \WHFast{\texttt{WHFast}}

\begin{document}

\title{Characterizing planetary systems with SPIRou: M-dwarf planet-search survey and the multiplanet systems GJ~876 and GJ~1148\footnote{Based on observations obtained at the Canada-France-Hawaii Telescope (CFHT) which is operated by the National Research Council (NRC) of Canada, the Institut National des Sciences de l'Univers of the Centre National de la Recherche Scientifique (CNRS) of France, and the University of Hawaii. The observations at the CFHT were performed with care and respect from the summit of Maunakea which is a significant cultural and historic site. } }

    \author{
C. Moutou\inst{1} \and
X. Delfosse\inst{2} \and
A.C. Petit\inst{3} \and
J.-F. Donati\inst{1} \and
E. Artigau\inst{4,5} \and
P. Fouqu\'e\inst{1} \and 
A. Carmona\inst{2} \and  
M. Ould-Elhkim\inst{1} \and 
L. Arnold\inst{6}  \and 
N.J. Cook\inst{4,5} \and 
C. Cadieux\inst{4,5} \and 
S. Bellotti\inst{1,12} \and
I. Boisse\inst{7} \and 
F. Bouchy\inst{8} \and
P. Charpentier\inst{1} \and
P. Cort\'es-Zuleta\inst{7} \and
R. Doyon\inst{4,5} \and 
G. H\'ebrard\inst{9} \and
E. Martioli\inst{10,9} \and 
J. Morin\inst{11}  \and
T. Vandal\inst{4,5}
}

\institute{
\inst{1} Univ. de Toulouse, CNRS, IRAP, 14 avenue Belin, 31400 Toulouse, France, \email{claire.moutou@irap.omp.eu} \\
\inst{2} Univ. Grenoble Alpes, CNRS, IPAG, 38000 Grenoble, France\\
\inst{3} Universit\'e C\^ote d’Azur, Laboratoire Lagrange, OCA, CNRS UMR 7293, Nice, France\\
\inst{4} Universit\'e de Montr\'eal, D\'epartement de Physique, IREX, Montr\'eal, QC, H3C 3J7, Canada\\
\inst{5} Trottier Institute for Research on Exoplanets, Université de Montréal, Département de Physique, C.P. 6128 Succ. Centre-ville, Montréal, QC H3C 3J7, Canada\\
\inst{6} Canada-France-Hawaii Telescope, CNRS, 96743 Kamuela, Hawaii, USA\\
\inst{7} Aix-Marseille   Université,   CNRS,   CNES,   LAM   (Laboratoire d’Astrophysique de Marseille),  Marseille, France \\
\inst{8} Geneva Observatory, University of Geneva, Chemin Pegasi 51, 1290 Versoix, Switzerland\\ 
\inst{9} Institut d'Astrophysique de Paris, UMR7095 CNRS, Universit\'e Pierre \& Marie Curie, 98 bis boulevard Arago, 75014 Paris, France \\
\inst{10} Laborat\'{o}rio Nacional de Astrof\'{i}sica, Rua Estados Unidos 154, 37504-364, Itajub\'{a} - MG, Brazil\\
\inst{11} Universit\'e de Montpellier, CNRS, LUPM,34095 Montpellier, France\\
\inst{12} Science Division, Directorate of Science, European Space Research and Technology Centre (ESA/ESTEC), Keplerlaan 1, 2201 AZ, Noordwijk, The Netherlands\\
}

\date{}

\abstract{
SPIRou is a near-infrared spectropolarimeter and a high-precision velocimeter. The SPIRou Legacy Survey collected data from February 2019 to June 2022, half of the time devoted to a blind search for exoplanets around nearby cool stars. The aim of this paper is to present this program and an overview of its properties, and to revisit the radial velocity (RV) data of two multiplanet systems, including new visits with SPIRou.
From SPIRou data, we can extract precise RVs using efficient telluric correction and line-by-line measurement techniques, and we can reconstruct stellar magnetic fields from the collection of polarized spectra using the Zeeman-Doppler imaging method. The stellar sample of our blind search in the solar neighborhood, the observing strategy, the RV noise estimates, chromatic behavior, and current limitations of SPIRou RV measurements on bright M dwarfs are described. 
In addition, SPIRou data over a 2.5-year time span allow us to revisit the known multiplanet systems GJ~876 and GJ~1148. For GJ~876, the new dynamical analysis including the four planets is consistent with previous models and confirms that this system is deep in the Laplace resonance and likely chaotic. The large-scale magnetic map of GJ~876 over two consecutive observing seasons is obtained and shows a dominant dipolar field with a polar strength of 30~G, which defines the magnetic environment in which the inner planet with a period of 1.94~d is embedded. For GJ~1148, we refine the known two-planet model. 
}

\keywords{ Planetary systems -- Techniques: radial velocities -- Instrumentation: spectrographs -- Stars: activity -- Stars: magnetic field -- Individual stars: Gl 876, Gl15A, GJ 1148}

\titlerunning{The SPIRou planet-search program }
\authorrunning{Moutou et al}
\maketitle

\section{Introduction}
\label{sec:Intro}
High-precision radial velocity (RV) observations have started in the optical domain and were recently extended to the near-infrared (NIR) spectral range, that is, beyond 950~nm. The relative performance and actual complementarity between optical and NIR stellar RV time series are a matter of active research, and they strongly depend on the nature and properties of the star. The technology for precise RVs in the optical domain has benefited from a longer development time, making instrumental and data-reduction techniques more mature than in the NIR. For instance, HARPS and ESPRESSO have inherited decades of instrumental and pipeline improvements \citep{mayor2020,cretignier2021,pepe2021}. The same applies, e.g., for EXPRES \citep{brewer2020}. In turn, RV instruments observing beyond 950~nm such as GIANO \citep{carleo2016}, CARMENES-NIR \citep{bauer2020}, IRD \citep{hirano2020}, HPF \citep{Metcalf2019}, SPIRou \citep{donati2020}, CRIRES+ \citep{dorn2023}, and NIRPS \citep{wildi2022} face new challenges, such as telluric correction, instrument stability at low temperatures, wavelength calibration across a new wide domain, and Hawaii detector persistence, which are still currently being characterized and dealt with. 
The NIR RVs, however, carry the excitement of discovering a rich wavelength domain populated by thousands of atomic lines and molecular bands. 
It is therefore of great interest and highly effective to investigate these NIR challenges in more detail to allow further explorations of stars and planets. 

In addition, the planet-finding surveys that trigger these RV developments are impacted by the magnetic activity of the stellar hosts, which is wavelength dependent \citep{reiners2013,baroch2020}. It is thus of prime interest to perform a planet search and characterization in the optical and NIR domains in order to establish achromatic planet signatures and to explore chromatic stellar signatures \citep{huelamo2008,reiners2010, mahmud2011, bailey2012, carmona2023}. While it may be challenging to reach an RV precision that is at the same level in the optical and NIR instruments within the same exposure time, instruments allowing both optical and NIR precision RVs simultaneously, such as CARMENES \citep{quirrenbach2014}, GIARPS \citep{claudi2016}, and HARPS/NIRPS \citep{wildi2022}, are in principle the best combination because stellar activity is also modulated in time.

Searches for planets around M stars have gained popularity since the population estimates obtained with Kepler and early RV surveys \citep{bonfils2013, dressing2015, gaidos2016}. Each M dwarf hosts at least two planets on average, with periods of 200 days, according to these occurrence studies. In addition, searching for Earth-like planets in the habitable zone of M stars is favored by the low mass of the star \citep[as in][]{bonfils2013,delfosse2013,anglada-escude2013}, both in terms of RV amplitude at a given planet mass and in terms of the period range in the habitable zone. The planets around M dwarfs in the solar neighborhood will also offer unique targets in the near future based on which the atmosphere of nontransiting exoplanets can be characterized with ELTs by separating the spectrum of the planets from that of their star through their changing relative Doppler shift \citep{snellen2015, lovis2017}. Planetary systems such as those of Proxima Cen \citep{anglada2016} and GJ 1002 \citep{suarez2023} already promise that these objectives can be reached. This ambitious objective of discovering nearby close-in telluric and/or habitable planets can be achieved from the ground only for the most favorable case: a planetary system with the lowest planet-to-star contrast and the largest angular separation. Several surveys of M stars are ongoing and are conducted for instance with HARPS \citep[][and references therein]{astudillo2017}, HARPS-N \citep{affer2016}, CARMENES \citep{sabotta2021}, or ESPRESSO \citep{Hojjatpanah2019,suarez2022}, and soon with NIRPS. In the mid-term, owing to the necessary partial redundancy of the target lists of these surveys, the combined data sets of RV measurements of nearby M dwarfs will offer many opportunities to characterize nearby planetary systems in reliable ways. 

The NIR high-RV precision spectropolarimeter and high-precision velocimeter SPIRou has been installed since 2018 at the Cassegrain focus of the 3.6\,m Canada-France Hawaii Telescope (CFHT). SPIRou has a resolution of 70\,000 over a nominal spectral range from 0.98 to 2.35\,$\mu$m and is stabilized at a milli-Kelvin temperature level in a vacuum chamber. It is fiber-fed with fluoride optical fibers through a science channel and a reference channel for simultaneous wavelength calibration. Calibration lamps allow the precise monitoring of the instrument, in particular, using the precision Fabry-P\'erot source \citep{hobson2020}. The science channel is composed of two fibers to convey the two polarimetric beams into the spectrograph. A detailed presentation of the instrumental design and a review of early measurement performances obtained with SPIRou are available in \citet{donati2020}.

In this paper, we present the planet-search survey around nearby M stars conducted with SPIRou between 2019 and 2022, in the framework of the SPIRou Legacy Survey. In Section~2 we present the survey definition: the stellar sample, observing strategy, and the data analysis method. The global observed properties of RV measurements are characterized in Section~3. Section~4 shows our analysis of a five-hour long sequence of SPIRou data of the bright M star GJ\,15 A. Section~5 further illustrates the SPIRou RV capability by describing the data analysis of two planetary systems orbiting the M stars GJ\,876 and GJ\,1148. Finally, section 6 contains conclusions and presents future work.

\section{SPIRou legacy survey planet-search program}
\label{sec:SLS}
A large observing program, the SPIRou legacy survey (SLS; PI J.-F. Donati\footnote{CFHT program IDs 19/20/21/22AP40 and 19/20/21BP40}), has started in February 2019, with the double objective of characterizing exoplanets of nearby stars and the magnetic field of their parent star. Within the 310 nights attributed to this program, the search for exoplanets around a sample including some of the most nearby and coolest stars of the solar neighborhood, the so-called planet-search (PS) program, accounts for half of the time. The second half of the time has been shared between the characterization of transiting planets and the study of young stellar systems \citep{donati2020}. The SLS data collection was completed in June 2022. A global description of the sample, the observing strategy, and the analysis pipeline of the SLS-PS (SPIRou Legacy Survey Planet-Search) survey is presented in this section. 

\begin{figure}
    \centering
\includegraphics[width=\hsize]{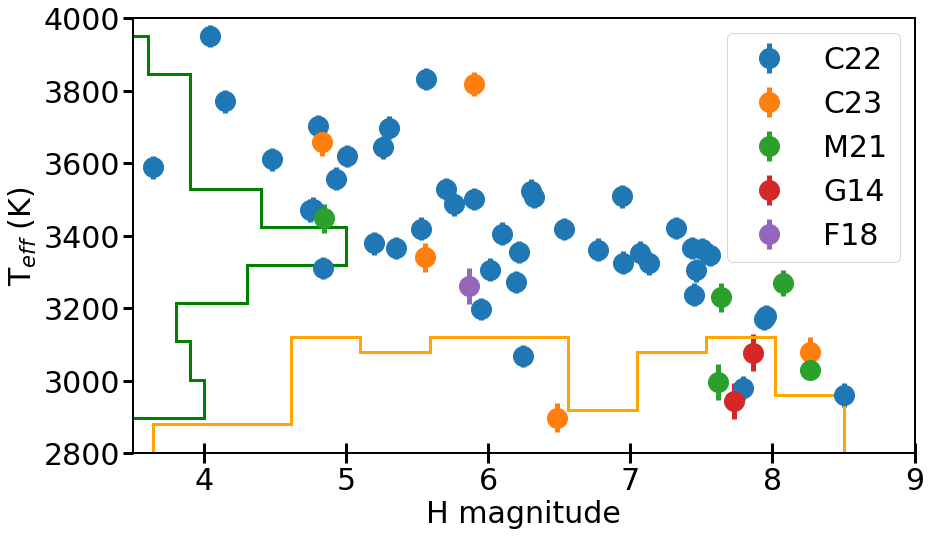}
\includegraphics[width=\hsize]{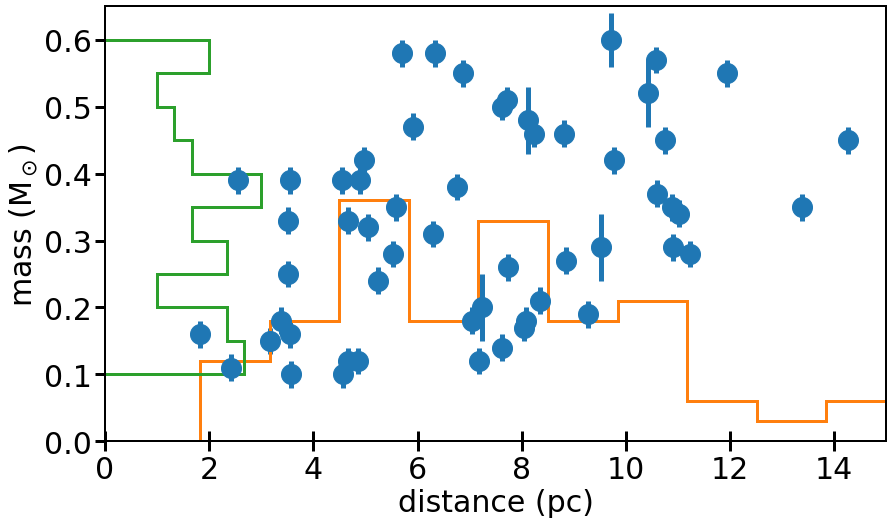}
\includegraphics[width=\hsize]{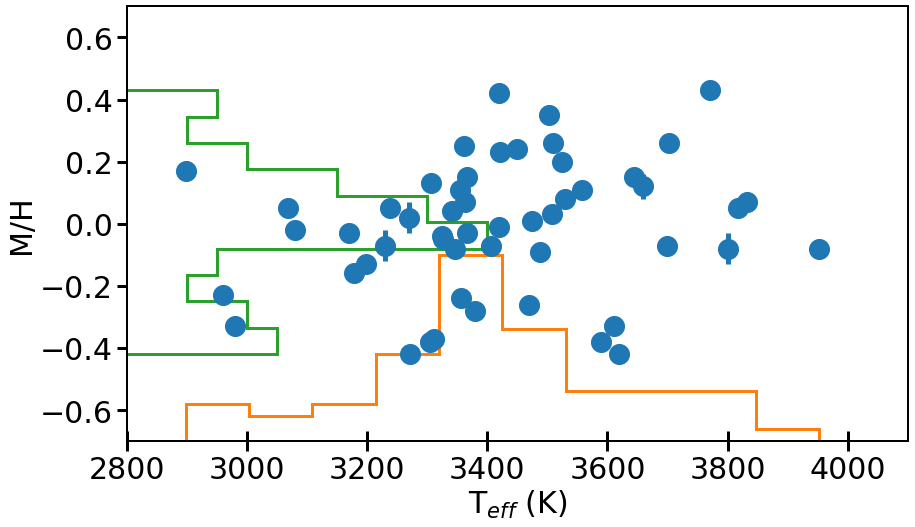}
\caption{Sample characteristics. Top: Distribution of effective temperatures as a function of $H$-band magnitudes in the stellar sample. The colors indicate the source of the temperature measurement: C22 \citep{Cristofari2022}, C23 \citep{cristofari2023}, M21 \citep{marfil2021}, G14 \citep{gaidos2014}, and F18 \citep{fouque2018}. Middle: Distribution of stellar masses and distances. Bottom: Distribution of metallicities and effective temperatures.}\label{fig:teffmh}
\end{figure}

\subsection{Stellar sample}
The planet-search program secured monitoring observations of 57 nearby M dwarfs; all stars in the sample are closer than 15\,pc, and most are closer than 10\,pc. These stars were chosen based on a preparatory analysis summarized in the SPIRou input catalog (SPIC) \citep{moutou2017, fouque2018}. SPIC was created from CFHT/ESPaDOnS observations of M stars acquired in the Coolsnap\footnote{CFHT programs 14BF13, 14BB07, 14BC27, 15AF04, 15AB02, 15BB07, 15BC21, 15BF13, 16AF25, 16BC27, 16BF27, and 17AC30.} program and archives. Criteria included the stellar distance, brightness, and mass \citep{fouque2018} and an activity merit function built from several activity indicators \citep{moutou2017}. Initially based on a sample of 100 stars \citep{cloutier2018}, the planet-search program had to be downsized to 57 stars in order to fit the reduced allocation in telescope time while maximizing the number of visits per star. This number includes a few stars that were observed in the framework of the calibration plan\footnote{program IDs 19/20/21/22AQ57 and 19/20/21BQ57} (see Section 2.2 for details). The distributions of effective temperature, $H$-band magnitude, mass, and metalicity of the observed stars are shown in Figure \ref{fig:teffmh}. This stellar sample represents about a quarter of the population of M stars \citep{reyle2021} within 10\,pc: there are 249 M dwarfs in this volume, 187 of which are visible from Maunakea; and 45 of our 57 stars lie within 10 pc. The range of observed $H$-band magnitudes is 3.64-8.50 and the temperature range is 2900-3850\,K, with a larger proportion of mid-M dwarfs with an effective temperature of about 3300-3400\,K. If 57\% of the M dwarfs within 10\,pc have spectral types M3 to M5 \citep{reyle2021}, our sample is biased toward these types and contains about 80\% of the same range. Effective temperatures, metallicities, and stellar masses were recently revised for 44 of the 57 stars \citep{Cristofari2022}. The other 13 stars were not included in this study either because only a few observations are available or because their  activity level is high. Their stellar parameters were retrieved from the literature (\citet{gaidos2014, fouque2018, marfil2021,cristofari2023}). The stellar masses of the sample range from 0.1 to 0.6\,\MS , and the metalicity ranges from [M/H] = -0.42 to +0.43\,dex \citep{Cristofari2022}.

Most chosen targets are stars known to have a low activity level. They either have a low H$_\alpha$ index, a long rotation period, or a low-level magnetic field, or all activity proxies are at moderate to low levels \citep{moutou2017}. There are a few exceptions, for which the objective was to test whether SPIRou data could identify the origin of the activity RV jitter, which is known to be large in the optical, as seen, for instance, for GJ~388 (AD Leo), GJ~3622, GJ~1111 (DX Cnc), GJ~1245B, GJ~873 (EV~Lac), GJ~406 (CN~Leo), and GJ~803 (AU~Mic) \citep{morin2008,morin2010,lannier2017}. The SPIRou observed properties of magnetic activity and RV data for AD~Leo are described in \citet{carmona2023} and \citet{bellotti2023}; for AU~Mic, they are described in \citet{martioli2020}, \citet{klein2021b}, and \citet{donati2023}; and for the more moderately active star GJ~205, they are described in \citet{cortes2023}. 

Sixteen planetary systems within the SLS-PS sample were discovered before the time of writing using optical RVs or transits, and seven of them were known before the start of the survey. They are listed in Table \ref{tab:knownplanets}. It is particularly interesting to study these system with the SPIRou SLS data set because differences or complementarity between the optical and NIR RVs is expected, especially regarding RV jitter induced by stellar activity. Two systems contain planets in transit (GJ~436 and AU Mic). As nearby planetary systems are the focus of future characterization studies, searching for additional companions and measuring the magnetic properties of the host stars are worth dedicating survey time to those known systems.

In addition, some of our targets are components of a stellar binary system. We give the angular separation of the components in 2016 to be compared to the diameter of the entrance fiber (1.28"). For some of them, we observed both M-type components of a stellar pair: GJ~15 A and B (angular separation 34"), and GJ~725~A and B (11.6"). At other times, only one known M-type component is suitable for SPIRou observations, and we observed this component of the system alone: GJ~752~A (75"), GJ~617~B (65"), GJ~169.1~A (10.3"), GJ~338~B (17"), GJ~412~A (32"), and GJ~1245~B (5.9"). Finally, we observed one SB1, TYC 3980-1081-1.

\begin{table}[]
    \centering
 \caption{Known planetary systems in the SLS-PS stellar sample ("q" means quadratic trend) and the most recent reference.}
    \begin{tabular}{lcl}
    \hline
    Star & N pl. & Reference \\ \hline
GJ~1002 & 2 & \citet{suarez2023} \\
GJ~15A  & 2 & \citet{pinamonti2018} \\
GJ~251  & 1 & \citet{stock2020}  \\
GJ~317  & 2 & \citet{feng2020}  \\
GJ~411  & 3 & \citet{hurt2022}\\
GJ~1148  & 2 & \citet{trifonov2020} \\
GJ~436  & 1 &  \citet{trifonov2018} \\
GJ~480  & 1+q & \citet{feng2020} \\
GJ~514  & 1 & \citet{damasso2022}\\
GJ~536 & 1 & \citet{suarez2017}\\
GJ~581  & 3 & \citet{trifonov2018} \\
GJ~687  & 2 & \citet{feng2020}  \\
GJ~752A  & 1 & \citet{Kaminski2018}\\
GJ~849  & 2 & \citet{feng2015}\\
GJ~876  & 4 & \citet{Millholland2018} \\
AU~Mic  & 2-4 & \citet{donati2023}  \\
\hline
\end{tabular}
    \label{tab:knownplanets}
\end{table}

\begin{figure}
    \centering
\includegraphics[width=\hsize]{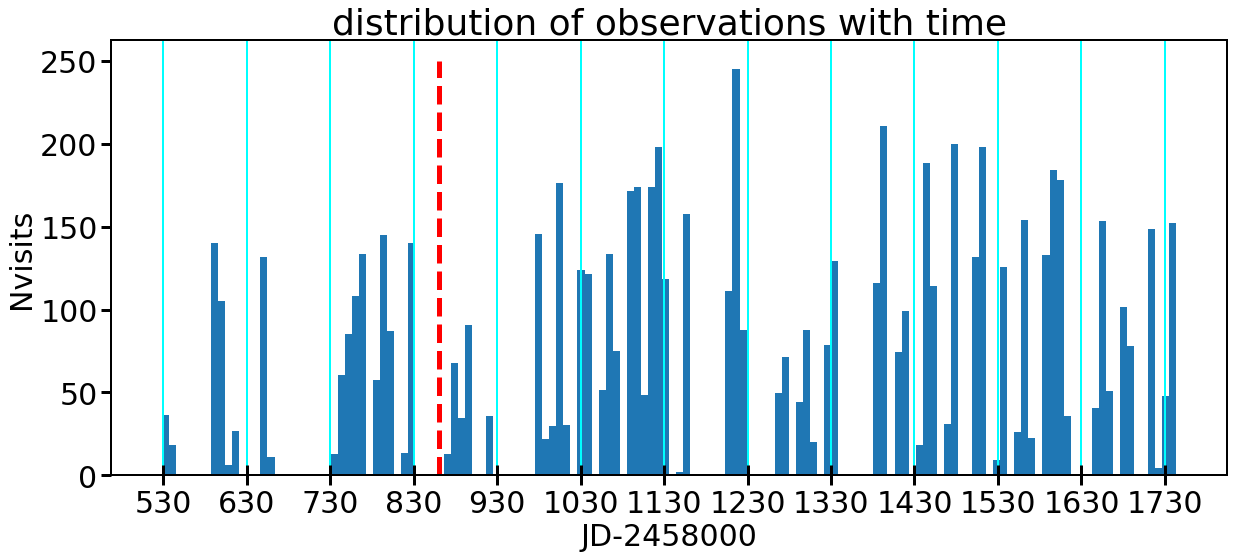}
\includegraphics[width=\hsize]{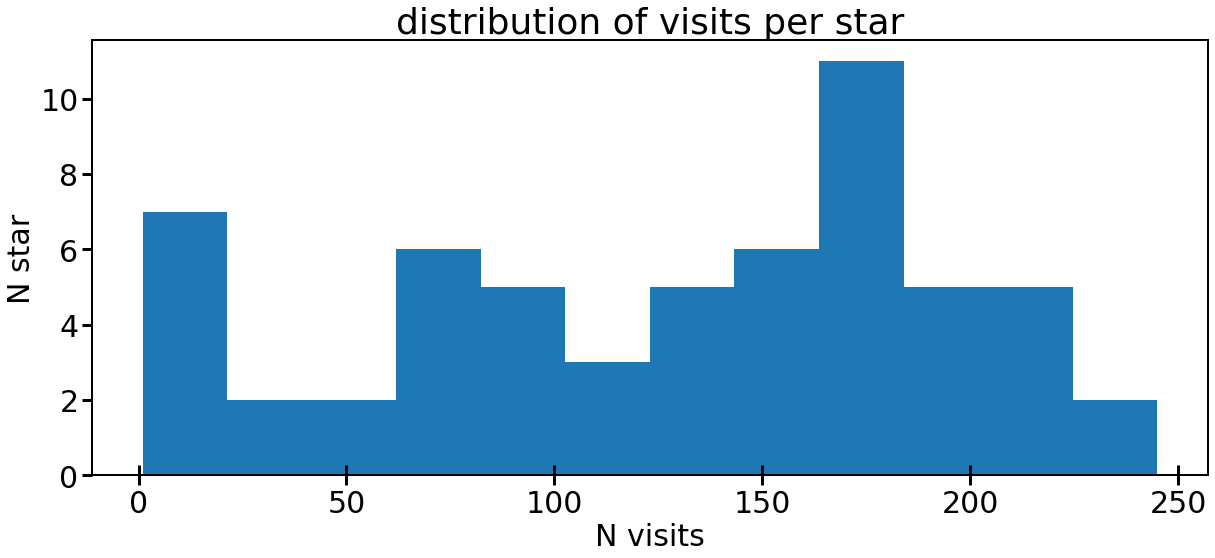}
    \caption{Sampling characteristics. Top: Distribution of observations with time, with weekly bins. The dashed red line shows the date of the maintenance thermal cycle of the spectrograph. Bottom: Distribution of visits per star of the sample.
    }\label{fig:log}
\end{figure}

\subsection{Observing strategy}
The survey benefits from the observatory calibration plan, which consists of daytime internal calibrations using lamps, a two-hour sequence that is performed in the afternoon and morning; and the telluric data base, consisting of hundreds of observations of hot fast-rotating stars. In addition, a few RV standards were observed each night according to the observatory selection, and these spectra are included in our survey. The CFHT RV standards are a mix of known planetary systems (GJ~436 and GJ~876) and stars of various spectral types known to be of low RV variability (GJ~846, GJ~382, GJ~514, and GJ~1286), and their SPIRou data are public at CADC\footnote{https://www.cadc-ccda.hia-iha.nrc-cnrc.gc.ca/en/search/}.

The monitoring rate, defined as the number of visits per observing season for each star, is highly important for any planet-search program, as was demonstrated with the detection of Proxima Cen\,b \citep{anglada2016}. Using SPIRou at the CFHT, exquisite observing conditions can be obtained during most of the year, but in turn, the instrument is not available full-time and is mounted during most of the bright-Moon periods. SPIRou runs typically last 10 to 15 nights. In total, the 310 nights of the SLS were spread over 494 SPIRou nights, covering a time span of 1100 nights. The relative fraction of the SLS programs in a given semester varies from 50 to 80\% of SPIRou nights from 2019 to 2022.
During the first year of data collection, a few events slowed the collection of data down: discontinued access to the observatory in July 2019, a technical shutdown of the telescope dome in August 2019, and operation disruptions due to the COVID-19 pandemic in April 2020. In addition, the observations were interrupted for a few weeks due to an instrumental thermal cycle in January 2020, caused by needed maintenance of the cold head (the dashed red line in Fig. \ref{fig:log}); this maintenance produced no notable offset in the measured RVs within the error bars. The monitoring rate has thus been quite low until the beginning of 2020, and then it increased significantly. 
Figure \ref{fig:log} (top) shows the distribution of the planet-search survey data collection as a function of time using weekly bins as a proxy for the monitoring rate. The epoch of the thermal cycling is also shown for reference.

The total observation time of the planet-search survey from February 2019 to June 2022 is 1044.8\,hours or 149.2 effective nights with the common conversion factor used for SPIRou of 7 effective observed hours per night. In total, 7153 visits were validated over a total of 7469 visits. A valid visit is defined as a complete Stokes V polarimetric sequence composed of four consecutive sub-exposures and a measured signal-to-noise ratio (S/N) per exposure of more than 70\% is the goal S/N.
The largest part of the program was executed in S/N mode, meaning that the exposure was terminated before the requested exposure time was reached when the preassigned goal S/N was obtained. The S/N of SPIRou exposures was monitored on the fly by summing up the flux on the science detector in an  area corresponding to the $H$ band of the stellar channel for each 5.572\,s up-the-ramp slice. The 15 brightest stars are not observed in S/N mode, but are rather observed at a fixed exposure time of 61.3s; they consequently have stronger S/N variations. Table \ref{tab:slsobs} lists the number of valid visits per star as well as their subexposure time, S/N ranges, and RV photon-noise uncertainty per subexposure (section 3.1). At the end of the survey, the number of validated visits per star was more than 200 visits for 7 stars, between 100 and 200 for 30 stars, and fewer than 100 visits for 20 stars. The distribution of visits per star is also shown in Figure \ref{fig:log} (bottom). 

Consisting of spectropolarimetric sequences in circular polarization, each visit simultaneously offers a precise RV measurement and a measurement of the longitudinal magnetic field $B_{l}$, as defined in \citet{donati1997}. Although these two quantities may not be related in a simple way \citep[e.g.,][]{hebrard2014, hebrard2016}, the magnetic field measurement may prove a valuable proxy for the activity jitter \citep[e.g.,][]{hebrard2016,martioli2020,klein2021b,Haywood2022,carmona2023} and a robust indicator of stellar rotational modulation \citep[e.g.,][]{klein2021a,cortes2023,fouque23}. 
Spectropolarimetry thus allows us to measure the rotation period of the star when the circular polarization signal is detected and at least quasi-periodic. This property was used to measure the rotation period of 27 stars in our sample \citep{fouque23} and to compare the values to the literature where other proxies were used for the rotation period.
Owing to the polarimetric sensitivity of SPIRou and observation strategy, the magnetic topology of most stars in this sample can be obtained with Zeeman-Doppler imaging \citep{donati2006} and with the field evolution over the 2.5 years of the survey. The first maps for a few stars in the SLS-PS sample are available in \citet{cortes2023} for the early-M star GJ~205 and in \citet{bellotti2023} for the mid-M star GJ~388. More maps are being analyzed for forthcoming publications.

Stellar RVs can be measured on each subexposure and can then be combined. Alternatively, the spectra can be combined before the RV is measured; this second method may show small timing differences (due to small variations in the overheads), however, and may finally be less accurate because of the precise correction for the Earth barycentric motion. RVs were therefore preferably measured on subexposures and were combined in the post-processing. All stellar RVs were corrected for the instrumental drift as measured in Fabry-P\'erot (FP) simultaneous spectra. A variable neutral density in the reference channel allowed us to tune the intensity of the FP so that the flux of the science and reference channels matched for $H$ magnitudes lower than $\sim$ 8.

As a side note, we calculated that the median seeing value of the whole SLS-PS data set is 0.72$\pm$0.20 arcsec, and the mean extinction is 0.07$\pm$0.20 magnitude. This illustrates the exquisite conditions at Maunakea Observatory.

\subsection{Pipeline and data analysis}

The SPIRou spectra were reduced with version 0.7.275 (March 2023) of the APERO pipeline \citep{cook2022}. APERO uses calibration frames to align the spectra with reference geometry and performs optimal extraction \citep{horne1986} of the science channels and the simultaneous Fabry-Pérot calibration channel. It also performs flat-fielding and corrects for thermal background and for the residual leak of the Fabry-P\'erot spectrum onto the science channel. Finally, based on the TAPAS model spectrum of the Earth atmosphere \citep{bertaux2014} and with the aid of a collected library of hot-star spectra obtained with SPIRou in various conditions of airmass and humidity, APERO also efficiently corrects for the telluric spectrum (\citet{artigau2014}, Artigau et al, in prep.). The final product is the 2D spectrum without the main instrumental contributions and corrected for telluric contamination \citep[see details in][]{cook2022}. A combination of the two science channels was used for RV measurements.

From these spectra, we then used the line-by-line (LBL\footnote{https://lbl.exoplanets.ca/}) method described in \citet{artigau2022} to determine the RVs. The LBL algorithm consists of measuring the RV contribution of very small chunks of spectra (lines) individually, with their proper weight. After obtaining line-by-line velocities, the algorithm rejects the lines giving outliers via a statistical analysis. The LBL algorithm was applied to both the stellar extracted spectrum and the Fabry-P\'erot spectrum in the reference channel, the latter giving a precise measurement of the instrumental drift. The final RV time series was corrected for this drift.
In addition to providing a mean RV value of the stellar spectrum, the LBL method also allowed us to select particular sets of lines, in order to filter out specific lines, for example. Thus, the \texttt{wapiti} software \citep{ouldelhkim2023} uses a PCA method on the individual line-by-line RV times series to determine the components that are not common to all lines (which may be due to insufficient telluric corrections or to differential activity effects) so as to retain only the common RV variations. LBL also allows checking for chromaticity in the RVs (see section \ref{sec:chrom}).

We then used the four exposures of the polarimetric sequence and the two independently extracted science channels to provide the polarized spectrum in the way described in \citet{donati1997}. Then, we used a least-square deconvolution (LSD; \citet{donati1997}) with a numerical mask that contained information about the sensitivity to the magnetic field of each identified line  (the Land\'e factor). We then obtained the Stokes I, V, and N LSD profiles of each sequence: the first is the mean intensity profile, the second is the circular polarization profile, and the third is a polarization check (null) to verify that the level of spurious polarization is negligible. The polarization pipeline Libre-ESpRIT was used in this paper. Details about this pipeline are available in \citet{donati2020}.

\subsection{Data }
\label{sec:data}
In this presentation and analysis of the SLS-PS survey data, we selected a subsample of the collected data and measurements to illustrate the capability of the survey while the bulk of the analysis is still taking place. 

Firstly, the RV photon noise estimates and the analysis of the short-term variations in polarimetric RV sequences are described together with relevant ancillary data when needed. 

Then, the five-hour-long sequence obtained on GJ~15A on October 7-8, 2020 is presented.
GJ~15A is an M2 dwarf star in a binary system. As a bright star located at only 3.56\,pc from the Sun, it has been extensively observed in optical RV surveys with ELODIE \citep{delfosse1998}, HIRES \citep{jenkins2009,howard2014}, HARPS-N \citep{pinamonti2018}, and CARMENES \citep{trifonov2018}. A few NIR RV data were also obtained with iSHELL \citep{cale2019}. GJ~15A is host to a 11.4-day planet \citep{howard2014}. Although questioned by \citet{trifonov2018} using HIRES and CARMENES data, this planet was later confirmed by \citet{pinamonti2018} with HARPS-N RV measurements. 
GJ~15A was observed with SPIRou from May 14, 2019, to January 25, 2022, with a total of 955 individual spectra. One hundred ninety validated polarimetric sequences were obtained on different nights, and an additional 48 consecutive sequences were obtained during a single technical night on October 7-8, 2020\footnote{CFHT program ID 20BP40}. All observations were made of four consecutive 61.3s exposures. In this paper, we mostly focus on the latter five-hour-long sequence. The whole data set that was collected on this star is only partly used for the comparison of noise properties, and its analysis regarding the planet signature will be presented in a forthcoming publication (Delfosse et al, in prep.). In general, RV time series of stable stars are not considered in this study and will be the focus of forthcoming articles.

Finally, data sets for two known multiplanet systems are included.
GJ\,876 is a quiet mid-M dwarf hosting a dynamically active four-planet system. The most prominent planetary signal was first discovered in 1998 \citep{delfosse1998}, and three other signals were successively discovered in the last 20 years \citep{marcy2001,rivera2005, rivera2010}. The three outer planets are trapped in a Laplace mean-motion resonance that was extensively modeled \citep[][and references therein]{Millholland2018}. The estimated RV jitter due to activity is 2\,m/s at most according to the literature.
GJ\,876 was observed 88 times with SPIRou in the framework of the observatory calibration plan\footnote{CFHT program IDs 19/20/21AQ57 and 19/20/21BQ57}, from June 2019 to October 2021 (time span of 865 days).

GJ\,1148 is a quiet mid-M dwarf hosting a two-planet system. The inner planet, an eccentric Saturn-like planet with a 41.4d period, was first discovered by \citet{haghighipour2010}. A second planet candidate with a similar projected mass and eccentricity as the inner planet was then found in a 532\,d period orbit \citep{butler2017} and confirmed with CARMENES data \citep{trifonov2018}. The system is dynamically stable, and the orbits are not in any commensurable resonant configuration \citep{trifonov2020}. The literature records no rotation-modulated activity and a residual RMS of 4 m/s \citep{trifonov2020}.
GJ~1148 was observed 101 times with SPIRou in the SLS-PS program from December 2019 to June 2022 (time span of 913 days) in search for additional planet signals and to characterize the magnetic activity.

For all data sets presented in this paper, we used RV data from the LBL analysis. For GJ~1148 data, we used the \texttt{wapiti}-corrected RVs \citep{ouldelhkim2023} that show fewer residual systematics than raw LBL data. This correction was not attempted on the short series of GJ~15A because this technique does not apply at short timescales. It was not selected for GJ~876 RVs either, where planetary signals are strong and dynamical effects are dominant, because of the high risk of residuals at some known planet signals, and because the interest is low due to the large weight of historical data compared to the limited SPIRou time series.

\section{Global radial velocity analysis}
\label{sec:analysis}
In this section, we present the RV noise measurements and their relations with the stellar properties, S/N, and wavelength. We also describe the RV dispersion or patterns within polarimetric sequences.

\subsection{ Radial velocity variability noise floor}
\begin{figure}
\includegraphics[width=0.9\hsize]{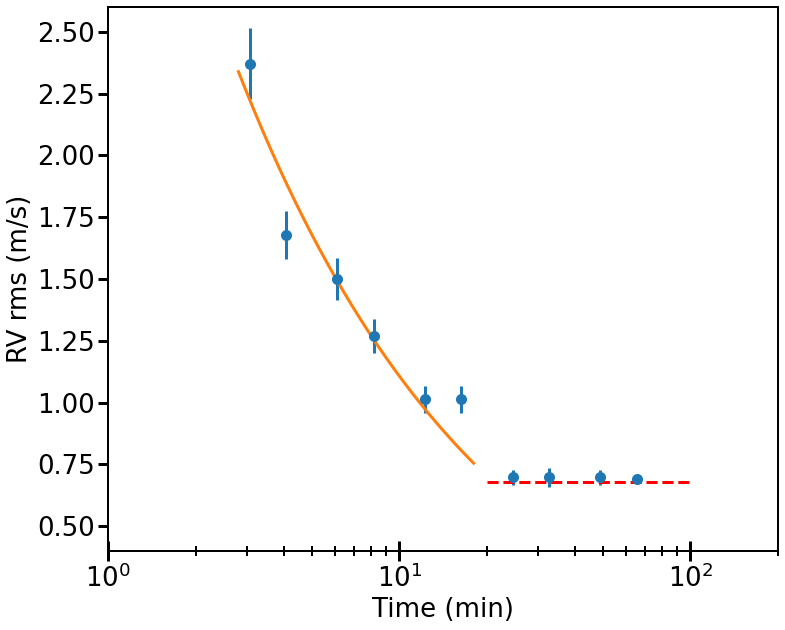}
\caption{RV dispersion during the five-hour sequence on GJ~15A as a function of the time bin. The orange curve represents the best-fit model in $t^{-1/2}$ up to the noise floor corresponding to an uncertainty of 0.7 m/s (dashed red line).}\label{rms_seq}
\end{figure}

In order to show the RV uncertainty noise floor for a given star, we studied the way in which the RV error decreases when more data on the same star were binned. We used the five-hour-long sequence on GJ~15A (192 spectra) presented in section \ref{sec:data} and successively binned the consecutive RVs of this series, regardless of the polarimetric sequences. Bins of 3, 4, 6, 8, 12, 16, 24, 32, 48, and 64 subsequent LBL RV time series (or total exposure times of 3~min to 64~min) were calculated. The error bars depict the dispersion in these bins. The result is shown in Figure \ref{rms_seq}. It exhibits the expected square-root of time dependence until a noise floor is reached. This noise floor is at 0.7 m/s for this particular star (nonrotationally broadened lines and effective temperature 3610~K), and it is achieved for an equivalent S/N of 1200 per extracted spectral bin in the $H$ band. Modal noise probably is the source of this noise floor.

\begin{figure}
    \centering
\includegraphics[width=0.9\hsize]{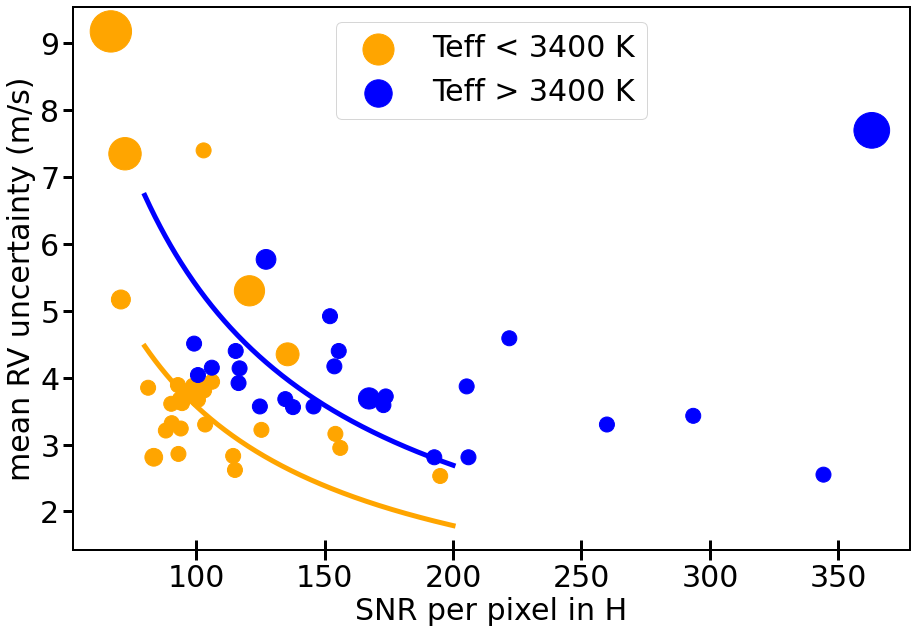}
\includegraphics[width=\hsize]{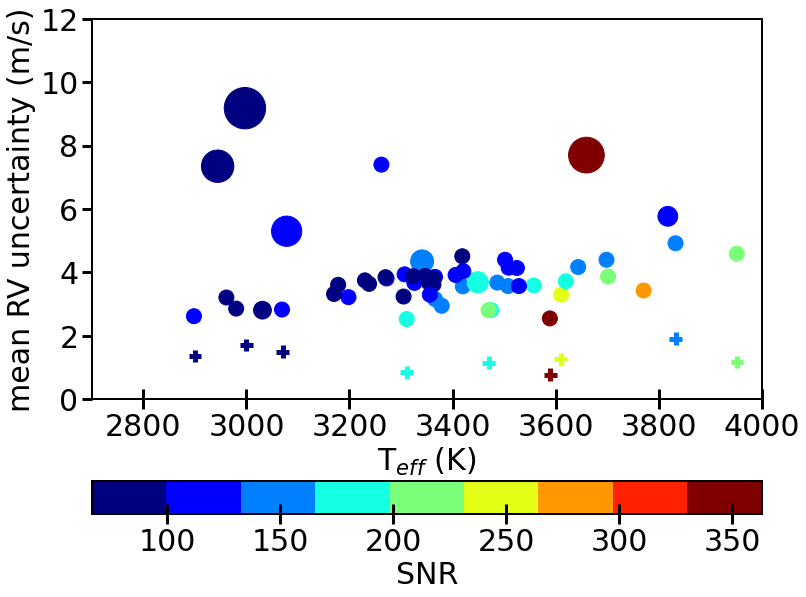}
    \caption{Distribution of the assessed mean RV uncertainty as a function of the mean S/N per pixel in the H band (top) and effective temperature (bottom). The symbol color shows the effective temperature of the star (S/N in the bottom plot), while its size is proportional to the projected rotational velocity. The lines in the top plot correspond to equations 1 and 2 (see text). The crosses show the values derived from \citet{reiners2020}.
    }\label{fig:phn}
\end{figure}

\subsection{Radial velocity photon-noise uncertainty and signal-to-noise ratio}

The effective RV uncertainty of a given spectrum depends on the achieved S/N, the RV content of the spectrum of the stars at the spectral resolution of the velocimeter, the projected rotational velocity, and other internal errors \citep{bouchy2001,figueira2016,artigau2018b,reiners2020}. 
As described in \citet{artigau2022}, the RV uncertainty in the LBL algorithm is obtained using a finite-mixture model, that is, by calculating the weighted mean in presence of outliers and its distribution (see Appendix B). 
Most of the program stars have an unresolved rotation profile; only a few objects have a short rotation period and a significant rotational broadening ranging from 3 to 10.5\,km/s, which strongly impacts their RV error. These stars are depicted with a larger symbol size in Fig. \ref{fig:phn}. We also  note that some stars have a noticeable Zeeman broadening of magnetically sensitive lines, which may also contribute to lower the RV accuracy \citep{cristofari2023,donati2023,carmona2023,bellotti2022}.

In Figure \ref{fig:phn} we show the mean RV photon-noise uncertainty for all stars in the SLS-PS sample as a function of the S/N, rotational velocity, and effective temperature. This average RV uncertainty per star and per subexposure is also listed in Table \ref{tab:slsobs}.

The tendency of the RV error is to increase for exposures of lower S/N values. As expected \citep{figueira2016}, at a given S/N and for slow rotators, the achieved RV precision in the NIR is better for the coolest stars by a factor of 1.5-2 than for the hotter stars in the sample. When the large-$v$sin$i$ profiles are filtered out and the RV error is fit as a function of S/N, global trends for the empirical RV uncertainty ($\sigma_{RV}$) appear in the S/N domain up to 200 per pixel, characterized by\begin{equation}
\indent    \sigma_{RV} \sim 550/S/N \rm \ m/s 
\end{equation}
for stars hotter than 3400\,K, and \begin{equation}
\indent  \sigma_{RV} \sim 350/S/N \rm \ m/s
\end{equation}
for cooler stars. This provides a mean accuracy of 5.5 and 3.5\,m/s for an S/N of 100. At higher S/N, the statistics are poor and a noise floor would depend on the spectral type, with values up to $\sim$ 2~m/s for stars hotter than 3400~K. 
The bulk of stars with highest-precision measurements with S/N values of about 100 or fewer (Figure \ref{fig:phn}, top plot) corresponds to the coolest stars in our sample.

\subsection{Short-term radial velocity stability}

\begin{figure}
\includegraphics[width=\hsize]{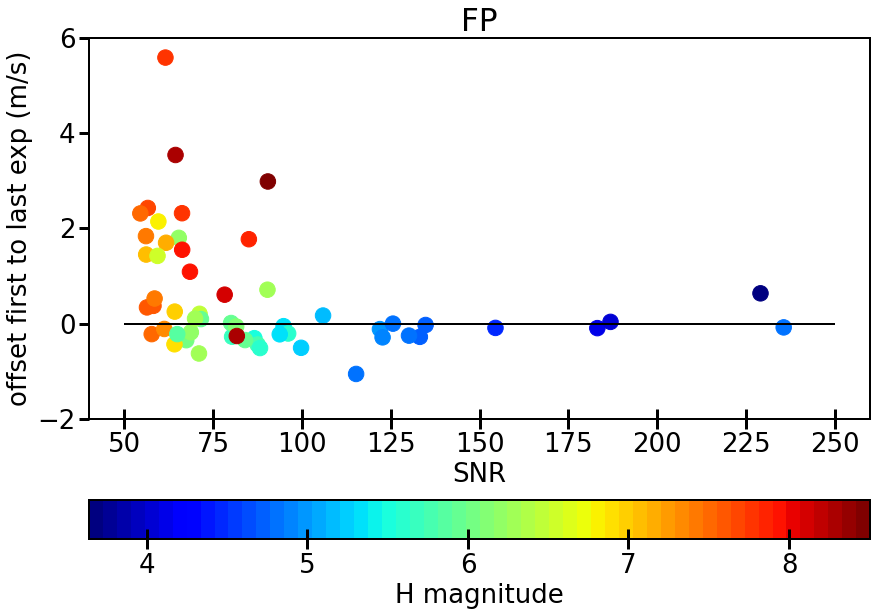}
\includegraphics[width=\hsize]{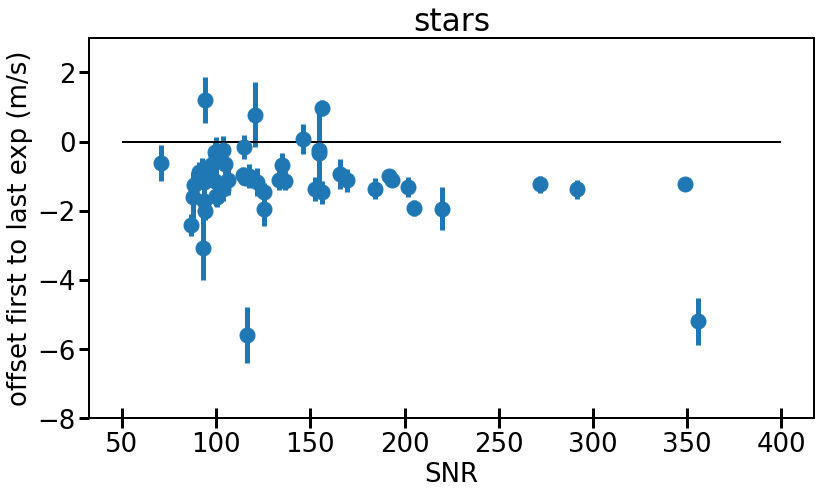}
\caption{Mean RV offset between the first and last exposure of the reference FP spectra (top) or the science channel (bottom) of a polarimetric sequence as a function of the S/N in the $H$ band for all sequences of the program. The black line indicates no offset. In the top plot, the symbols are colored with the respective magnitude of the star that is observed in the science channel. 
}\label{fig:pattern_seq}
\end{figure}

Searching for possible systematic patterns in the RVs within a polarimetric sequence, we used all 7420 spectra and grouped them into polarimetric sequences per star. Then we calculatedthe mean difference between the first and last exposure of the sequence   for each star of the sample, using raw (not corrected for drift) stellar RVs. The RVs tend to decrease with time over the sequences, that is, within 5 to 30 minutes. Fifty-five percent of the stars depart from zero at 3$\sigma$ , and all these biases are negative (i.e., a decreasing RV trend with time). The mean negative offset between the first and last exposure is about -1.2 m/s. This is shown in Figure \ref{fig:pattern_seq} for the whole sample, except for the most active stars where the measurement error is too large (GJ~1111, GJ~1245B, and GJ~1256 are not included, nor is TYC 3980-1081-1, which is an SB1). 
Some stars (AU Mic and GJ~581) unexpectedly have a much larger offset of about 6 m/s than
the bulk of stars. 

The effect is also seen in the reference channel, and we therefore performed this separately on both channels. We looked at the RVs measured in the reference channel during the night for SLS-PS exposures. The flux is this channel was adjusted to the stellar magnitude in order to have similar flux in both the science and reference channels. This match is valid up to an $H$ magnitude of about 8. As shown in Figure \ref{fig:pattern_seq} (bottom), the offset is reversed with the FP lamp, and the dispersion at lower S/N is strong. The offset on the FP amounts to 2-3 m/s (sometimes 4-6 m/s) at the lowest S/N and finally disappears at S/N values higher than 100. The pattern on the science channel tends to be more dispersed at any S/N values, probably because the RV content varies from one star to the next.

Further investigations of this pattern will be conducted to evaluate whether a correction is required or if a noise term should be added. The origin of this small effect could be in the detector readout/persistence, in the optics, in a residual contamination between channels, in a residual linkage from one channel to the next, or in the pipeline, it may depend on the night history,  or be a mix of several of those effects, and it requires additional testing. On the other hand, due to the great stability of the instrument, we exclude that the spectrograph drift is the cause of these offsets: as a typical example, the five-hour sequence on GJ~15A has a drift standard deviation of 0.34~m/s (see Table 3). The global instrument drift is therefore negligible in sequences of a few minutes.
In normal operations, the stellar RVs corrected for the simultaneous FP calibration is used. This offset is a constant for a given star, and the impact on drift-corrected RV data is negligible. It should, however, be taken into account whether common trends are estimated, or if nightly instrumental drifts are monitored.
In order to further identify the role of the polarimeter in these offsets, we made some additional tests that we present below.

\subsection{Radial velocity dependence on the polarimeter configuration}
\label{section:rhomb}
\begin{figure}
\includegraphics[width=1.1\hsize]{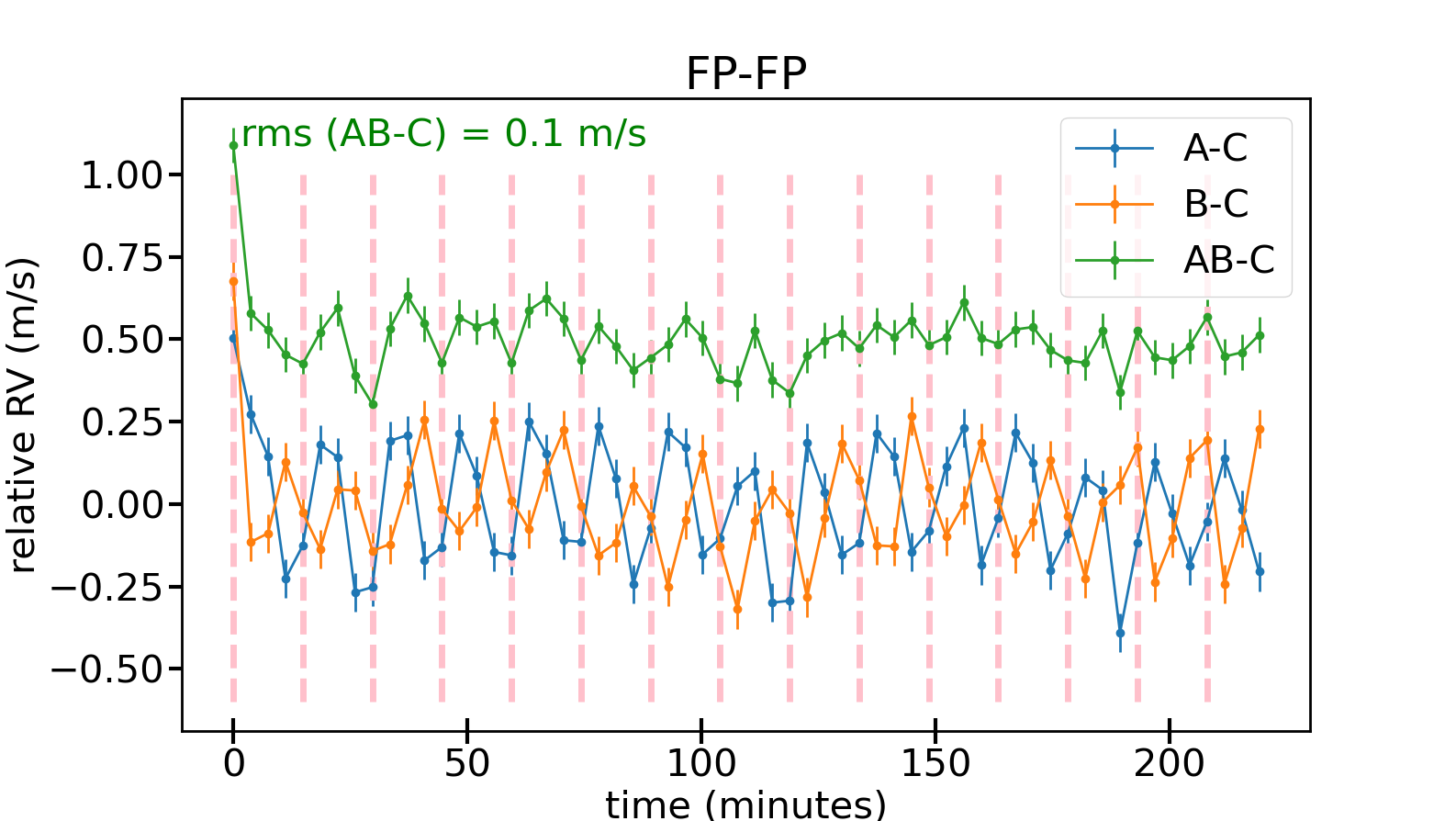}
\caption{Relative RVs of a sequence of Fabry-P\'erot exposures in the science channels (A and B in blue and orange, respectively) as a function of time relative to the first exposure and corrected for the simultaneous calibration channel (C) RVs. The green curve shows the combined science-reference relative RV drift, shifted by 0.5 m/s for clarity. The vertical pink lines separate different polarimetric sequences.}\label{fp_rv}
\end{figure}

In order to evaluate the impact of the polarimeter on the measured RVs, we first used long series of Fabry-P\'erot (FP) exposures in the polarimetric mode. Light from the FP etalon follows the same path as the star from the calibration wheel to the spectrograph, crossing the rhombs and the Wollaston prism in the science channel. This science channel is thus separated into two independent channels, named A and B. Channel C is the reference channel used for simultaneous calibration. Each channel can be extracted separately by the pipeline. When the conditions of science polarimetric observations are simulated by rotating the rhombs, FP spectra are recorded on both the science fibers and the calibration fiber (this latter goes directly from the calibration unit to the spectrograph). The relative RVs show a distinctive zigzag pattern of $\sim$ 0.25 m/s amplitude related to the polarimeter configuration in the individual channels A and B separately during this 3.7h continuous series, with dispersion values of 18 and 17 cm/s, respectively, relative to the reference channel. However, they mostly compensate for each other, and the total science channel (AB), used for precision-RV work, shows RVs that are stable at a level of 10 cm/s. The FP time series is shown in Figure \ref{fp_rv}: separate and combined science channels relative to the reference channel. Before correction by the reference, the standard deviation of the separate science and reference channels are 0.37, 0.41, 0.38, and 0.35 m/s for channels AB, A, B, and C, respectively.

\begin{figure}
\includegraphics[width=\hsize]{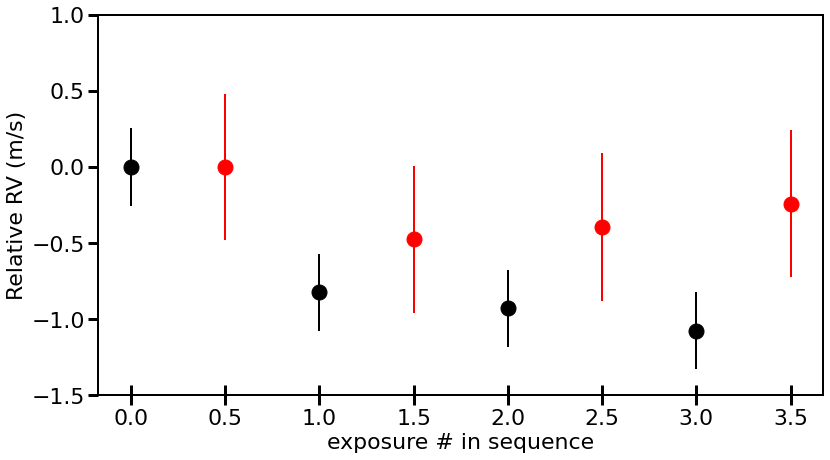}
\caption{Relative RVs of GJ~15A in the science channel as a function of time relative to the first subexposure, representing the position of the polarimeter. In red we plot the long series of consecutive sequences collected in October 2020, and in black we show the average of all other data. The data points are shifted by 0.5 in X for clarity. }\label{polar_rv}
\end{figure}

To complement the test performed for the Fabry-P\'erot polarized spectra, we looked at the five-hour-long polarimetric data on  GJ~15A. Folding of all 48 four-minute polarimetric sequences in time shows that the stellar RVs are stable in all four configurations of the polarimeter with a mean first-to-last exposure difference of $-0.24\pm0.48$ m/s. 
For comparison, the mean relative difference in nightly polarimetric sequences of GJ~15A is superimposed in  Figure \ref{polar_rv}. The first-to-last exposure RV offset is $-1.08\pm0.25$ m/s.

The slightly larger offsets seen when the star is observed during a single sequence on a night may be due to persistence/flux effects on the detector, as illustrated in section 3.2 and expected from H4RG detectors \citep{artigau2018}. This is suggested by the fact that the offset becomes negligible when the same star is observed for a long time, hence with constant illumination.

\begin{figure}
    \centering
\includegraphics[width=0.95\hsize]{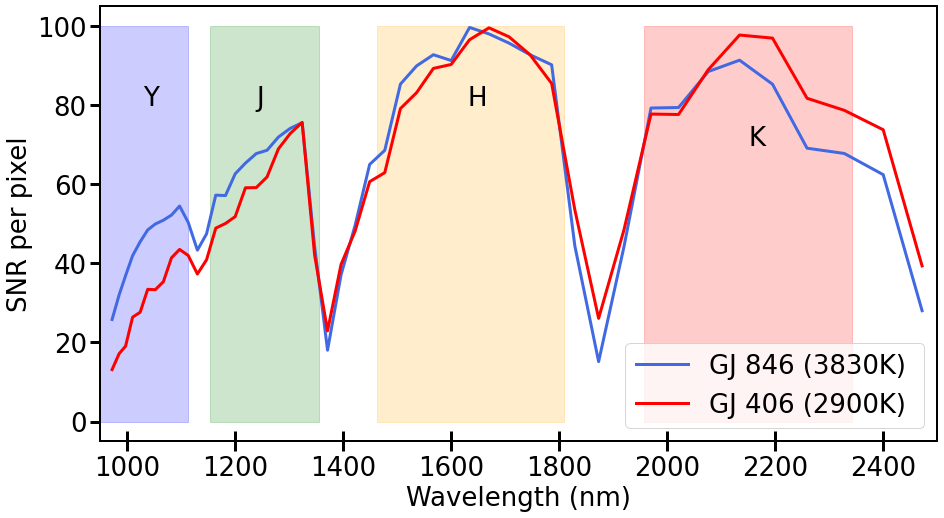}
\includegraphics[width=0.95\hsize]{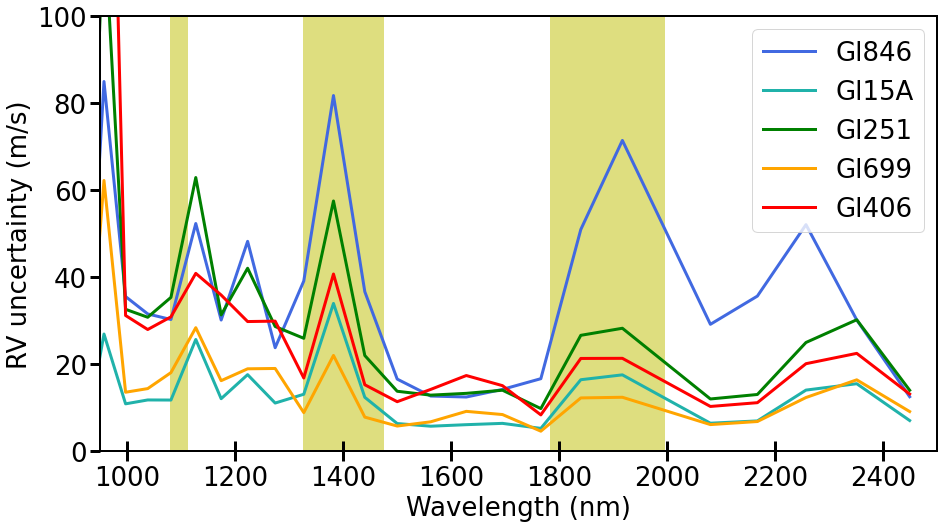}
\includegraphics[width=\hsize]{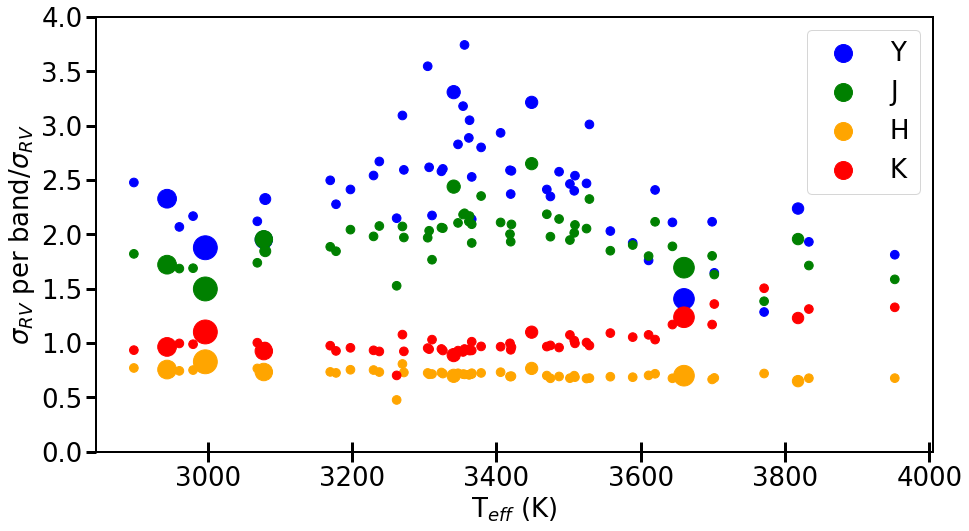}
    \caption{Chromatic behavior of the SPIRou data. Top: Relative S/N as a function of wavelength for the hottest and coolest stars in the sample. The S/N values have been normalized at 1650 nm. Middle: RV uncertainty for a selection of program stars with decreasing effective temperatures from top to bottom in the legend plotted along the spectrum. The yellow areas show the bands with maximum telluric absorption. The RV uncertainty has been normalized to an S/N of 150 at 1650 nm. Bottom: Ratio of the median RV error in a single band and the median RV error in the whole spectrum as a function of effective temperature for the whole data set. Larger symbols represent the few stars with a resolved projected rotational velocity. 
    }\label{fig:chrom}
\end{figure}

\subsection{Radial velocity uncertainty dependence on wavelength}
\label{sec:chrom}
The LBL analysis \citep{artigau2022} allows us to group lines together in spectral chunks and study the behavior of the RV uncertainty as a function of the stellar spectral type. We checked the chromatic behavior of the photon noise for several stars with high S/N and various effective temperatures. In Fig. \ref{fig:chrom} (middle), each chunk encompasses about two spectral orders. This shows that the RV noise is lowest in the $H$ band for all stars, as found on Barnard's star by \citet{artigau2018b}. The hottest stars in the sample, such as GJ~846, have large uncertainties in the $K$ band compared to cooler stars. The uncertainty also increases in all stars at the location of large Earth atmospheric absorption bands, but it is nevertheless notable that some measurements are possible in these ranges despite the reduced access to the stellar spectra. 
As shown in Figure \ref{fig:chrom} (bottom), where all stars are represented, there is a small decreasing trend of the RV uncertainty in the $H$ band toward hotter stars and an opposite behavior in the $K$ band. For the $Y$ and $J$ bands, the tendency is for the $3200-3500$~K stars to have larger RV uncertainties in these bands than for stars below 3200~K or beyond 3500~K. This is consistent with the lower S/N in the $Y$ and $J$ bands (Figure \ref{fig:chrom}, top), although it is not clear what the relative roles of RV content, modal noise, and S/N can be, globally, in these observed trends. The active star AU Mic (large symbol near 3650~K) has relatively low uncertainties in all bands owing to its brightness and observed high S/N values, and the noise is lower in the $Y$ than $J$ band, in contrast to other stars in the same temperature range. The definition of the $YJHK$ bands is shown in Figure \ref{fig:chrom} (top).

\subsection{Long-term radial velocity stability}
For longer timescales, the difficulty of assessing RV stability in the stellar channel lies with the ubiquitous existence of astrophysical signals (planet companions and/or stellar activity). For Barnard's star, the current measurement of dispersion of SLS-PS data is 2.59~m/s \citep{artigau2022}. A general way is to search for common trends in the time domain (e.g., for SOPHIE; \citet{courcol2015} or HIRES; \citet{talor2019}), as RV variations can be expected from seasonal thermo-elastic deformations of the spectrograph or from residual errors of the wavelength solution. With telluric contamination, it is also possible to search for common trends in a parameter space related to the relative velocity between the stellar and telluric spectra. Another way is to trust the data to drive the search for systematic effects, and correct for them, as was explored by \citet{ouldelhkim2023}. These two alternate approaches are the subject of ongoing work  (Artigau et al, in prep., and Ould-Ehlkim et al in prep.) and will not be detailed further here.

\section{Five-hour sequence on GJ 15A}
\label{sec:5hseq}
\subsection{Detailed short-term behavior of the GJ~15A sequence}
\begin{figure}
\includegraphics[width=\hsize]{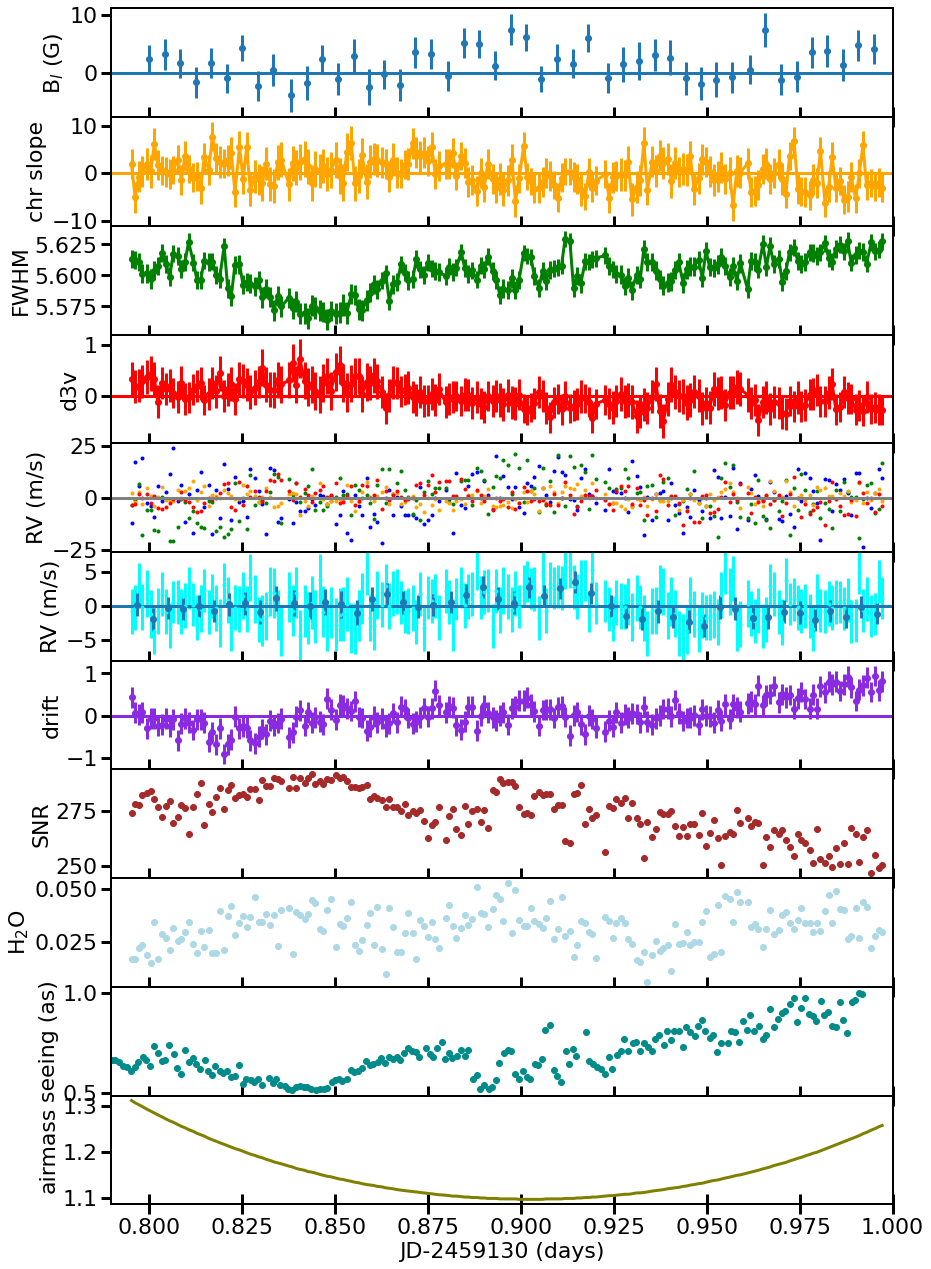}
\caption{Evolution of the measured parameters during the 5h sequence on GJ~15A. From bottom to top: Airmass, image quality, amount of atmospheric water, S/N in the H band, RV FP drift, RVs (cyan for individual spectra, blue for polarimetric sequences), RVs per band (blue, green, orange, and red for $Y$, $J$, $H$, and $K$ bands, respectively), $d3v$, FWHM, chromatic slope, and longitudinal magnetic field.}\label{fig:seq}
\end{figure}

The five-hour long sequence on GJ~15A obtained as an engineering program on October 7, 2020,  was used to explore the instrumental stability and the way in which the stellar properties measured by SPIRou behave at these hourly timescales and with respect to ancillary data.
We show in Figure \ref{fig:seq} the behavior of the whole data set: airmass, image quality, an estimate of absorbing telluric water, S/N, RV FP drift, total LBL RVs, RVs per band, $d3v$,  dLW, chromatic slope, and longitudinal field. The image quality was estimated during the sequence on the guiding image. The differential line width (dLW) is the coefficient associated with the Taylor expansion of the line profiles, as introduced by \citet{Zechmeister2018} and included in the LBL method \citep{artigau2022}, which behaves like the departure from the mean stellar FWHM. The parameter $d3v$ is the third coefficient of the Taylor expansion of the line profiles, representing a proxy for higher-order distortions of the profile and behaving like a bisector span. The chromatic slope over the SPIRou wide spectral range is another byproduct of the LBL analysis and contains the chromatic dependence of the RVs on wavelength, whereas the combined RVs is the (achromatic) velocity at a reference wavelength of 1600~nm \citep{artigau2022}. It is similar to what is measured in the visible across CARMENES orders, as described in \citet{Zechmeister2018}.

The notable features in these data series are i) a decrease in stellar width at the beginning of the series (time index near 0.85 d) that has no counterpart at the same phase in RV, but is reflected in a larger RV error at the same time; ii) a slight trend in the S/N, FWHM, and $d3v$ over the whole sequence; and iii) a rather flat behavior of the RVs, particularly during the FWHM dip, with a slightly decreasing trend in the second half of the sequence. It is unclear whether this decrease can be related to the seemingly more unstable S/N or to a slight change on the stellar surface that is also visible in the $d3v$ parameter and the chromatic slope (with a phase lag). The S/N behavior is clearly correlated with the variation in image quality, and there is a suspicious correlation between seeing and FWHM. The cause for this correlation is unclear. It is further investigated in the appendix. The relation between FWHM, RV, S/N, seeing, and RV uncertainty is also worth exploring further on a large data set. 
Table \ref{tab:gl15a} lists the measured standard deviation and mean error of all parameters in the sequence. The dispersion is smaller than the mean error for $d3v$ and all RVs, for B$_l$ and chromatic slope, and larger only for FWHM and the absolute drift. It is possible that our mean errors (a byproduct of the finite-mixture model of LBL) are currently overestimated. The mean value of the water content estimated from telluric correction is three times larger than average during this series than in the whole data set of this star, and the dispersion is four times lower: a stable, but humid night.

\begin{table}[H]
\caption{Standard deviations of the measured parameters during the 5h sequence on GJ~15A.}\begin{tabular}{lcc}\hline
Parameter                        & Std deviation & Mean error\\
\hline
B$_l$ (Gauss)                    & 2.62  & 2.75\\
Chrom slope (m/s/$\mu$m)         & 2.93 & 3.22\\
FWHM (m/s)                 & 14.7 & 7.3 \\
$d3v$ (relative)                 & 0.22 & 0.32\\
RVs ($Y$ band, m/s)                & 9.45 & 10.59\\
RVs ($J$ band, m/s)                & 9.39 & 11.09\\
RVs ($H$ band, m/s)                & 3.38 & 4.63\\
RVs ($K$ band, m/s)                & 4.89 & 7.36\\
RVs ($YJHK$, indiv, m/s)           & 2.23 & 3.31\\
RVs ($YJHK$, polar, m/s)           & 1.42 & 1.70\\
FP drift (m/s)                   & 0.34 & 0.25 \\
S/N in $H$ band                    & 11   & - \\
H$_2$0 content (mm)              & 0.01 & 0.03 \\
\hline
\end{tabular}
\label{tab:gl15a}
\end{table}

\begin{figure}
\includegraphics[width=\hsize]{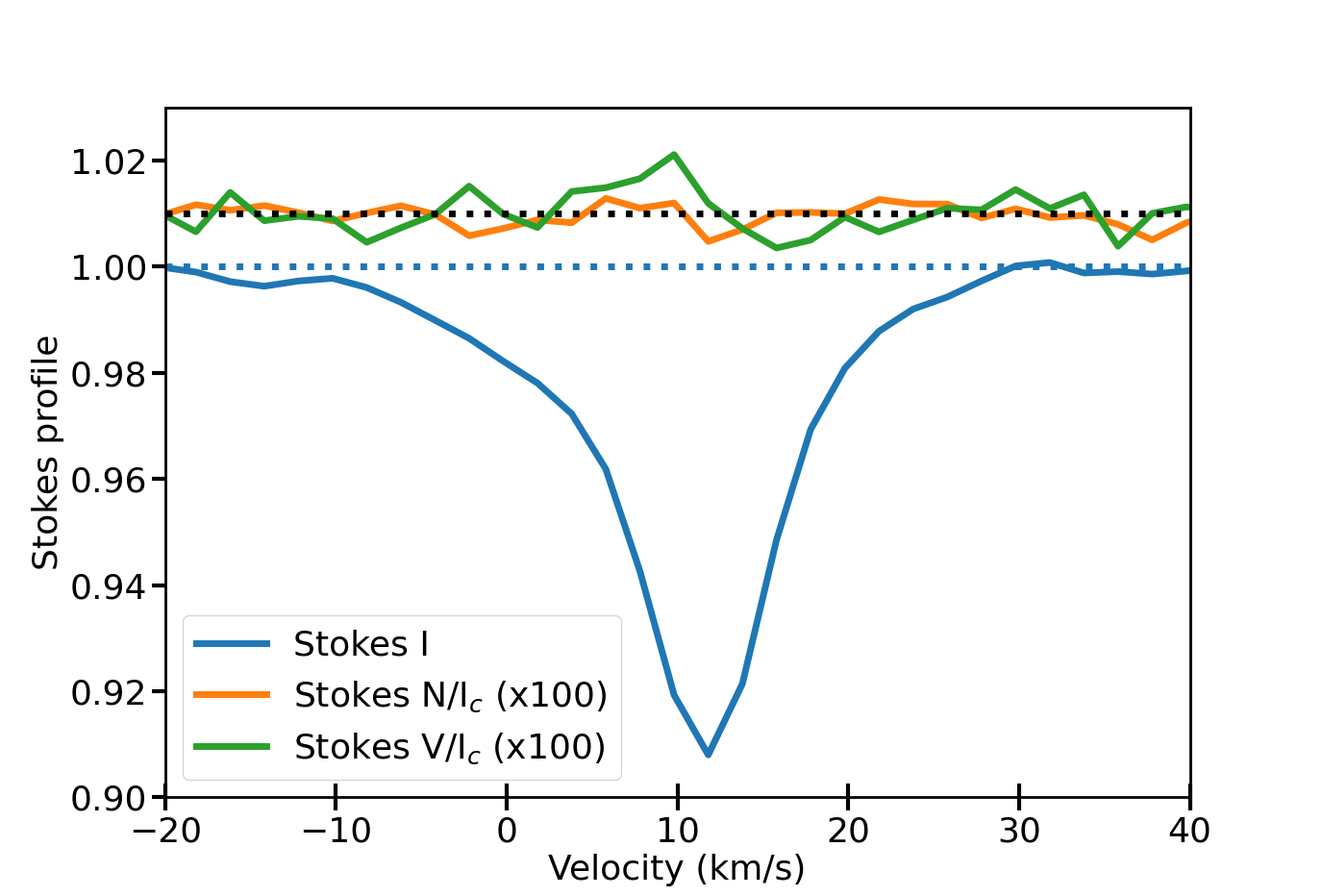}
\caption{Mean Stokes profiles of GJ~15A during the 5h long sequence. Stokes V and N profiles are multiplied by a factor of 100 and are shifted for more clarity. 
}\label{fig:stokes}
\end{figure}

\subsection{Spectropolarimetry}
The five-hour-long sequence on GJ~15A was conducted in the circular polarimetric mode so that we can observe how the polarimetric signal builds up and how the noise decreases with time. As shown in Table \ref{tab:gl15a} and Fig. \ref{fig:seq}, the longitudinal field is constant over the sequence, with a mean value of 1.6$\pm$2.6 G. Figure \ref{fig:stokes} shows the final Stokes I, V/I$_c$ , and N/I$_c$ profiles after they were summed over the whole sequence (where I$_c$ is the continuum intensity level). The standard deviation in the median null profile N/I$_c$ is 2.2 10$^{-5}$, illustrating the extreme polarimetric sensitivity of SPIRou. The precision of 10 ppm on the reflected solar spectrum was also measured at an S/N per pixel of 2000 \citep{donati2020}.

\section{SPIRou radial velocity observations of known planets around GJ~876 and GJ~1148}
\label{sec:planets}

\begin{figure}
    \includegraphics[width=0.9\linewidth]{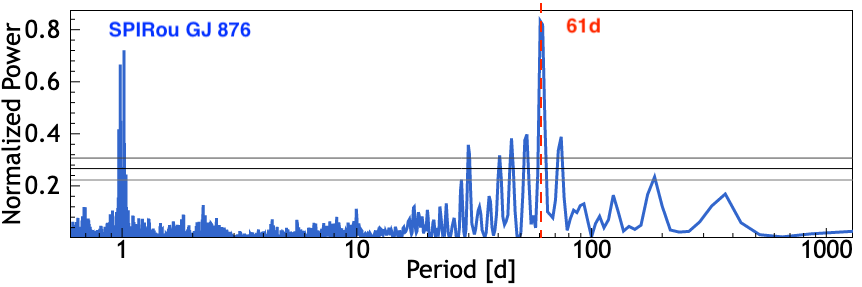}
    \includegraphics[width=0.9\linewidth]{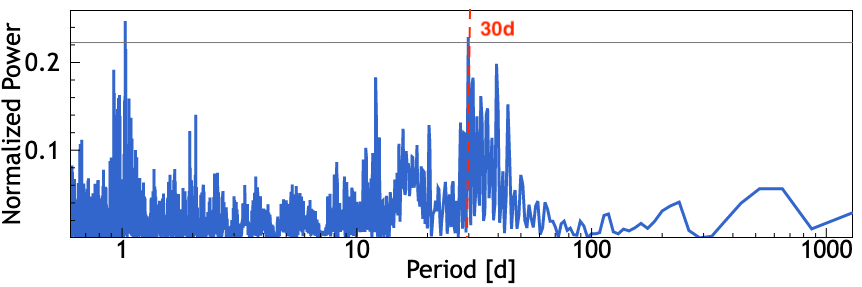}
    \includegraphics[width=0.9\linewidth]{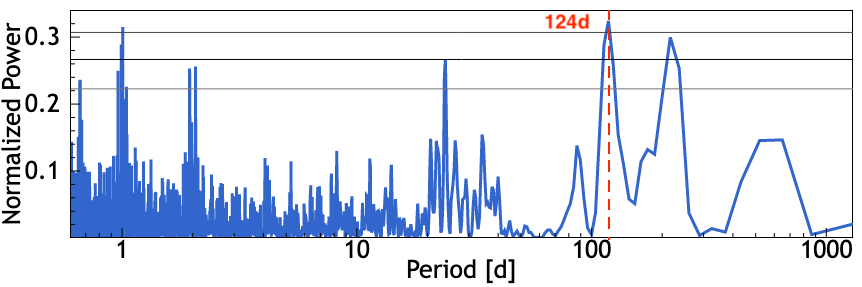}
    \includegraphics[width=0.9\linewidth]{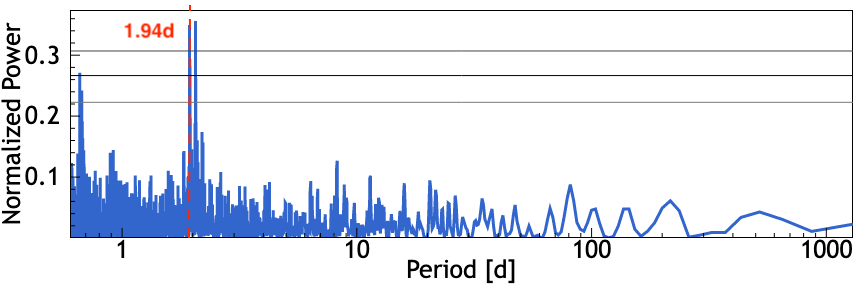}
    \caption{Sequential periodograms of SPIRou RV data of GJ~876 showing each planet signal, starting with raw data and after successively removing the previous signal. From top to bottom: 61d peak, 30d peak, 124d peak, and the 1.94d peak and its 1d alias. The horizontal lines correspond to 10\%, 1\%, and 0.1\% FAP values, respectively, from bottom to top. \label{fig:GJ876spirouP}}
\end{figure}

\begin{figure*}
    \includegraphics[width=\linewidth]{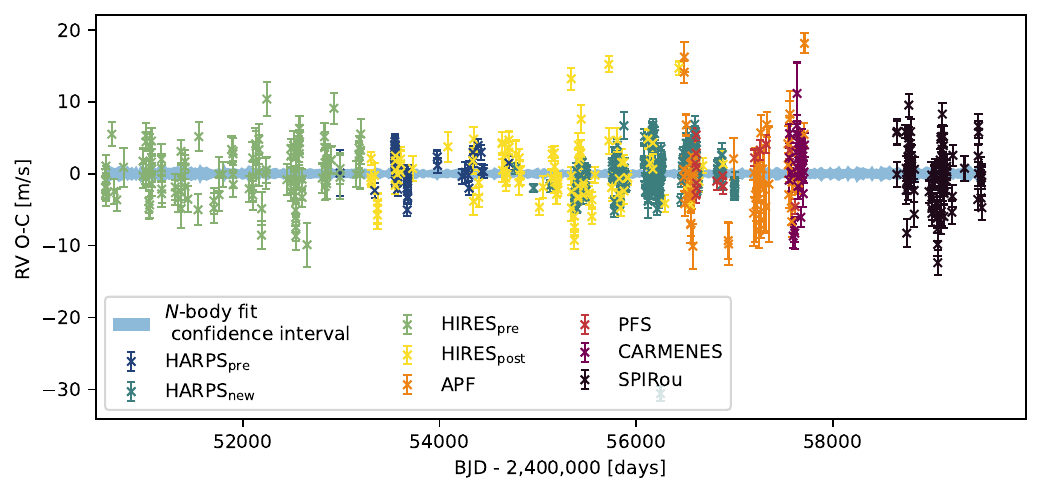}
    \caption{RV residuals obtained by subtracting the instrument offsets and the $N$-body fit the observed data. 
    The blue shaded region corresponds to the RMS of the modeled RVs of 200 randomly selected configurations from the fit posterior.\label{fig:RVresiduals}}
\end{figure*}

\subsection{Multiplanet system hosted by GJ\,876}
GJ\,876 hosts four known planets, three of which are in a Laplace resonance (planet b at 61\,d, planet c at 30\,d, and planet e at 124\,d periods) and a fourth one, planet d, in the inner part of the system with a 1.94\,d orbit.
The 88 SPIRou RVs allow the detection of all four planets, with a final standard deviation of the residuals of 3.3\,m/s, when the dynamical interactions over the three-year span are ignored and a four-Keplerian model is adjusted to the SPIRou data. 
Figure \ref{fig:GJ876spirouP} shows the Lomb-Scargle periodograms of SPIRou data alone, when planet signals are sequentially removed, using DACE \citep{delisle2016}. Because of the mean-motion resonance and suboptimal data sampling (SPIRou mostly is a bright-time instrument; see the window function shown in Figure \ref{fig:gj876p}), the 30d signal is barely detected, while the lower-amplitude signals of planets d and e are easily detected, with $log_{\rm FAP}$ of -4.13 and -3.45, respectively.

When using all historical RV data obtained on this system, however, the dynamic interactions between the planets cannot be neglected. In the following, we present the results of our dynamical analysis of the four-planet signal and the results after adding the new RV data to data obtained with HIRES (June 1997 to August 2014), HARPS (December 2003 to December 2014), both retrieved in \citet{Millholland2018}, and CARMENES (June to November 2016 from \citet{trifonov2018}). 

The resonant interactions between planets b, c, and d induce a rapid evolution of the Keplerian orbits. These mutual perturbations were used to provide strong orbital constraints through $N$-body fits \citep[e.g.,][]{Laughlin2001,Rivera2001,rivera2010,correia2010}.
In particular, this approach allows lifting the degeneracy between the inclination of the system and the planet masses.
The system is almost coplanar and inclined by \(\sim55^\circ\) with respect to the plane of the sky.

Following the approach from the most recent works on GJ~876 \citep[][hereafter M18]{Nelson2016,Millholland2018}, we performed a full $N$-body fit assuming a coplanar system in inclination with respect to the line of sight. We included the RVs already used by \citetalias{Millholland2018} and the 28 measurements made by CARMENES \citep{trifonov2018} and the 88 SPIRou visits.
Unlike \citetalias{Millholland2018}, we did not use a Gaussian process to model the residual signal and instead kept a simpler white-noise jitter model.
We cannot expect the stellar activity to have the same behavior in the infrared as in the visible, which would significantly complicate the modeling while bringing a limited improvement to the fit.
This choice is further justified by the limited amplitude (2.2\,m/s) of the activity jitter found by \citetalias{Millholland2018}.
Finally, in this simulation, we used the stellar mass of $M_s=0.33$\ M$_\odot$ from \citet{Cristofari2022}, which is 10\% lower than the value used in \citetalias{Millholland2018}.
More details on the used method are given in Appendix \ref{app:GJ876}.

\begin{table*}
    \caption{Best-fit parameters from the full four-planet coplanar $N$-body fit$^a$. \label{tab:GJ876}}
\begin{center} \begin{tabular}{ l c  c  c  c }  \hline\hline Parameter & Planet d & Planet c & Planet b & Planet e \\ \hline
$P$ (days) & ${1.9377904}^{+0.0000064}_{-0.0000073}$ & ${30.1039}^{+0.0069}_{-0.0066}$ & ${61.1035}^{+0.0062}_{-0.0069}$ & ${123.55}^{+1.0}_{-0.59}$ \\[1mm]
$a$ (AU) & ${0.021020525}^{+0.000000047}_{-0.000000053}$ & ${0.130874}^{+0.00002}_{-0.000019}$ & ${0.209805}^{+0.000014}_{-0.000016}$ & ${0.3355}^{+0.0019}_{-0.0011}$ \\[1mm]
$K$ ($\mathrm{m\ s^{-1}}$) & ${6.0}^{+0.19}_{-0.19}$ & ${87.48}^{+0.22}_{-0.22}$ & ${212.01}^{+0.21}_{-0.21}$ & ${3.49}^{+0.23}_{-0.23}$ \\[1mm]
$m$ ($\mathrm{M}_\oplus$) & ${6.68}^{+0.22}_{-0.22}$ & ${235.3}^{+2.5}_{-2.4}$ & ${749.2}^{+8.7}_{-8.3}$ & ${16.0}^{+1.0}_{-1.0}$ \\[1mm]
$e$ & ${0.035}^{+0.033}_{-0.024}$ & ${0.257}^{+0.0018}_{-0.0019}$ & ${0.0296}^{+0.003}_{-0.0013}$ & ${0.0545}^{+0.0069}_{-0.022}$ \\[1mm]
$\sqrt{e}\cos\omega$ & ${-0.06}^{+0.13}_{-0.11}$ & ${0.3229}^{+0.0055}_{-0.0074}$ & ${0.127}^{+0.018}_{-0.012}$ & ${-0.096}^{+0.08}_{-0.072}$ \\[1mm]
$\sqrt{e}\sin\omega$ & ${0.11}^{+0.11}_{-0.15}$ & ${0.3911}^{+0.0057}_{-0.005}$ & ${0.114}^{+0.014}_{-0.013}$ & ${-0.185}^{+0.19}_{-0.045}$ \\[1mm]
$\omega$ (deg) & ${120.0}^{+70.0}_{-56.0}$ & ${50.45}^{+1.0}_{-0.81}$ & ${41.7}^{+6.5}_{-6.2}$ & ${240.0}^{+23.0}_{-50.0}$ \\[1mm]
$M+\omega$ (deg) & ${217.4}^{+4.1}_{-4.3}$ & ${343.7}^{+0.39}_{-0.39}$ & ${16.08}^{+0.21}_{-0.19}$ & ${207.3}^{+5.4}_{-8.3}$ \\[1mm]
$i$ (deg) & \multicolumn{4}{c}{${56.7}^{+1.0}_{-0.99}$} \\ [1mm]
$\Omega$ & \multicolumn{4}{c}{0 (fixed)} \\
\hline
\multicolumn{5}{c}{Instrumental Jitters and Offsets$^b$  ($\mathrm{m\ s^{-1}}$)} \\
\hline
  $\sigma_{\mathrm{jit,HIRES_{pre}}}$ & ${2.8}^{+0.3}_{-0.28}$ &  & $\gamma_{\mathrm{HIRES_{pre}}}$ & ${25.41}^{+0.33}_{-0.33}$\\[1mm] 
   $\sigma_{\mathrm{jit,HIRES_{post}}}$ & ${3.81}^{+0.33}_{-0.29}$ &  & $\gamma_{\mathrm{HIRES_{post}}}$ & ${28.59}^{+0.43}_{-0.42}$\\[1mm]
   $\sigma_{\mathrm{jit,HARPS_{pre}}}$ & ${1.74}^{+0.26}_{-0.23}$ &  & $\gamma_{\mathrm{HARPS_{pre}}}$ & ${94.07}^{+0.39}_{-0.38}$\\[1mm] 
   $\sigma_{\mathrm{jit,HARPS_{post}}}$ & ${3.07}^{+0.18}_{-0.17}$ &  & $\gamma_{\mathrm{HARPS_{post}}}$ & ${-4.58}^{+0.23}_{-0.24}$\\[1mm] 
   $\sigma_{\mathrm{jit,APF}}$ & ${5.17}^{+0.62}_{-0.54}$ &  & $\gamma_{\mathrm{APF}}$ & ${81.56}^{+0.75}_{-0.75}$\\ [1mm]
   $\sigma_{\mathrm{jit,PFS}}$ & ${2.56}^{+0.74}_{-0.54}$ &  & $\gamma_{\mathrm{PFS}}$ & ${-19.75}^{+0.81}_{-0.79}$\\[1mm] 
   $\sigma_{\mathrm{jit,CARMENES}}$ & ${2.65}^{+0.64}_{-0.53}$ &  & $\gamma_{\mathrm{CARMENES}}$ & ${-287.5}^{+0.62}_{-0.61}$\\[1mm] 
   $\sigma_{\mathrm{jit,SPIRou}}$ & ${3.52}^{+0.36}_{-0.32}$ &  & $\gamma_{\mathrm{SPIRou}}$ & ${-1306.15}^{+0.45}_{-0.45}$\\[1mm] 
\hline
 \end{tabular}
     \footnotesize
     \item $^a$ The error bars correspond to the 68\% confidence interval
     \item $^b$ $\gamma$ values are relative to the stellar mean systemic velocity
\end{center} 
\end{table*}

The RV residuals obtained after subtracting the instrument offsets and the dynamical fit to the data are shown in Figure~\ref{fig:RVresiduals}.
The SPIRou measurements have a fitted jitter of 3.52\,m/s, to be compared to the jitter from optical data from HARPS (1.74 to 3.04\,m/s), HIRES (2.8 and 3.81\,m/s), and CARMENES (2.65\,m/s).
The root mean squared (RMS) of the residuals for the SPIRou data set is $3.78\,\mathrm{ms^{-1}}$.
When we instead take the best-fit value from \citetalias{Millholland2018} and model the RVs without fitting to the data, the RMS of the SPIRou data set residuals is $4.33\,\mathrm{ms^{-1}}$.
As expected for a planetary system that is very well covered by RV surveys, the newly obtained data are consistent with previous orbital characterizations.

The parameter estimates from the coplanar four-planet dynamical fit are reported in Table~\ref{tab:GJ876}.
As expected from the different adopted stellar mass estimates, the planet masses and inclination are shifted from the fit performed by \citetalias{Millholland2018}.
Nevertheless, nearly all our parameters fall in the $1\sigma$ range of the estimation from \citetalias{Millholland2018}.
The main exception is the quantity $\sqrt{e_\mathrm{e}}\sin{\omega_\mathrm{e}}$, which \citetalias{Millholland2018} estimated to be $0.04^{+0.07}_{-0.12}$ , whereas we find a value of  $-0.185^{+0.19}_{-0.045}$.
Both our fit and \citetalias{Millholland2018} find that this quantity has a bimodal distribution, with both reported values being explored by the MCMC, but with a different preference in each case. 
We attribute the discrepancy between the two fits to our choice of not using a Gaussian process to model the stellar activity.
When we ran our model on the data analyzed by \citetalias{Millholland2018}, we obtained a result similar to our complete analysis.
Because of the strong resonant interactions between planets c, b, and e, the dynamical fit is mostly sensitive to the resonant component of the eccentricities, as in the case of a transit-timing variation (TTV) fit \citep{Hadden2016,Petit2020}.
The difference in the eccentricity of planet e between the two fits has a marginal effect on the magnitude of the resonant variable.

The SPIRou RV observations of GJ\,876 are consistent with previously determined orbital configurations of the system.
The main differences with the orbital determination of \citetalias{Millholland2018} are mostly driven by the differences in stellar activity modeling.
Regarding the dynamical properties of the system, we confirm that the system is deep in resonance and is likely chaotic. 

Figure \ref{fig:gj876p} shows the periodogram of the SPIRou RV residuals and the SPIRou ancillary data of the LBL method, which may be sensitive to the magnetic activity signal. The dLW parameter, a proxy of the mean FWHM of the spectral lines we used, mainly shows power at the measured rotation period of 83.7d (see next section). The period is slightly displaced but compatible within 2$\sigma$ with the period seen in the longitudinal magnetic field parameter, B$_\ell$. This is expected because B$_\ell$ and dLW may be sensitive to different types of active regions. A peak at the rotation period may also be noted in the RV residuals, but it has a low significance. In turn, neither the chromatic slope nor the $d3v$ parameter (proxy for the bisector span) shows significant power at this period. We also verified that external values, such as the seeing or S/N, show no significant peak in the periodogram, and that seeing and B$_\ell$ are not correlated, as expected.

\subsection{Large-scale magnetic topology of GJ~876}

\begin{figure}
    \centering
\includegraphics[width=0.9\hsize]{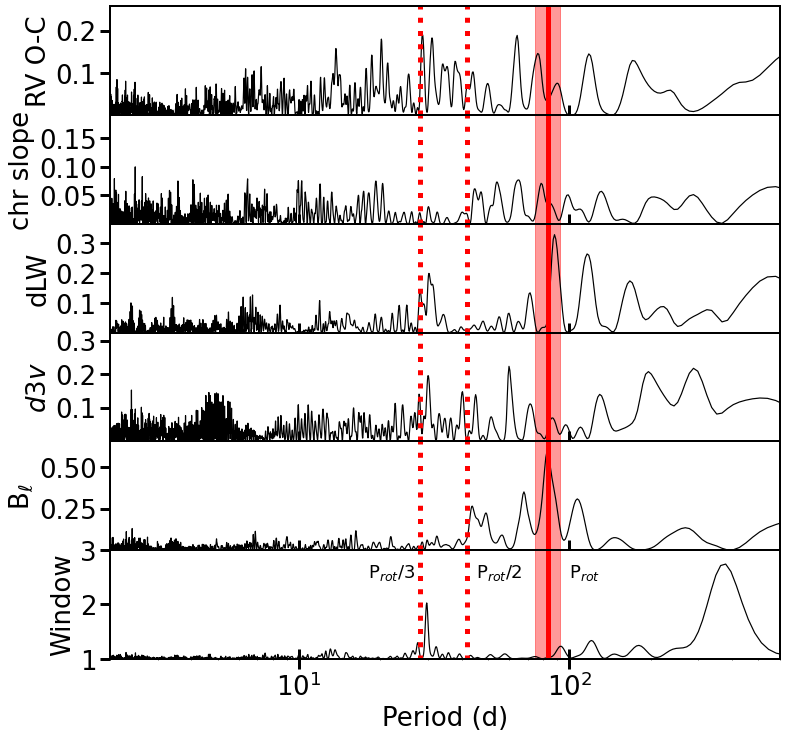}
    \caption{Periodograms of SPIRou ancillary LBL data (three middle rows), spectral window (bottom row), B$_\ell$, and RV residuals (top row) for GJ~876. The red zone shows the rotation period (Donati et al, in prep) and its 3$\sigma$ range. The first two harmonics of the rotation period are indicated as dotted red lines.}\label{fig:gj876p}
\end{figure}

\begin{figure*}
    \centering
\includegraphics[width=0.9\hsize]{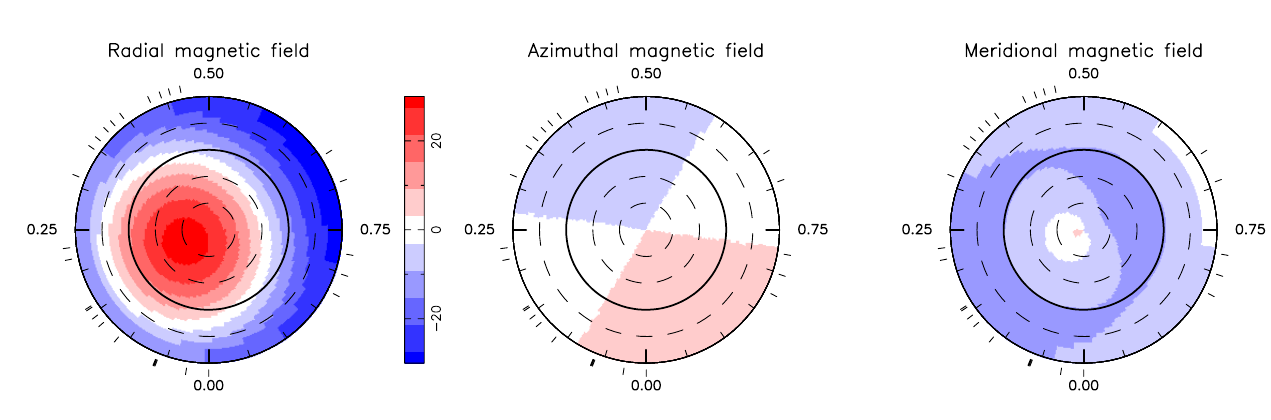}
\includegraphics[width=0.9\hsize]{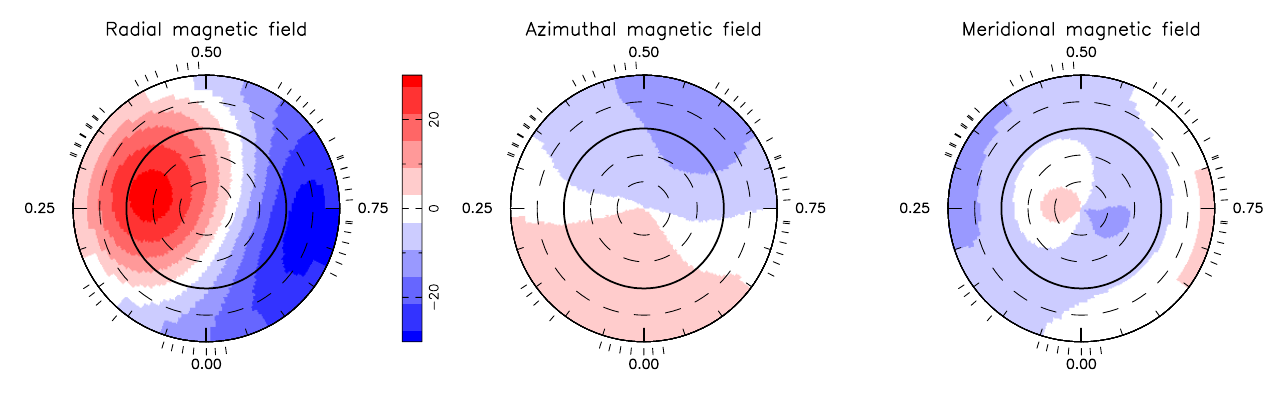}
\caption{Reconstructed maps of the large-scale field of GJ\,876 in 2019 (top) and 2020 (bottom). The left, middle and right columns represent the radial, azimuthal, and meridional field components in spherical coordinates. The color bar is expressed in Gauss. The maps are shown in a flattened polar projection down to latitude --60\degr\, with the  pole at the center and the equator depicted as a bold line. Outer ticks indicate phases of observations, assuming a rotation period of 83.7~d.}
\label{fig:gj876b}
\end{figure*}

In parallel to the RV modeling, we analyzed the circularly polarized (Stokes $V$) data of the slowly rotating host star GJ~876 in order to map the large-scale magnetic topology using Zeeman-Doppler imaging \citep[ZDI;][]{donati2006}.  We applied a least-squares deconvolution \citep[LSD;][]{donati1997} to all recorded Stokes V spectra using an M3V mask as described in Donati et al 2023 (in prep).  We then applied ZDI to our 28 LSD Stokes V profiles collected in 2019 and repeated the same operation on the 50 LSD profiles collected in 2020. The remaining ten measurements in 2021 do not sample the whole period and were thus left out of this analysis. We used the rotation period of 83.7$\pm$2.9~d found by modeling the longitudinal field variations (Donati et al 2023, in prep, and periodogram in Figure~\ref{fig:gj876p}) and assumed an inclination of the rotation axis of the star with respect to the line of sight $i=60\degr$.  The large-scale field of GJ~876 is mainly poloidal at both epochs and can be approximated by a 30~G dipole, tilted at 30\degr\ and 60\degr\ with respect to the rotation axis in 2019 and 2020, respectively (see Fig.~\ref{fig:gj876b}). The magnetic flux is damped with distance as $(a/R_s)^{-3}$. As the planet distance is 13.5 times the stellar radius, the scaling factor is 2460 on the magnetic flux. With a mean dipole between 20 and 40 G, modulated by latitude, inclination, and phase, one estimates an order of magnitude of 8-16~mG for the magnetic field at the orbit of the inner planet.

\begin{figure}
    \includegraphics[width=0.9\linewidth]{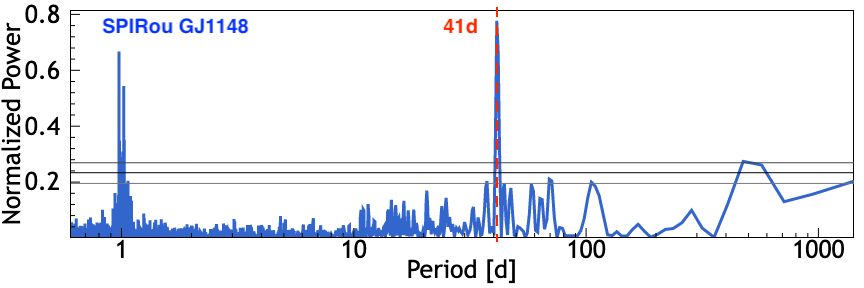}
    \includegraphics[width=0.9\linewidth]{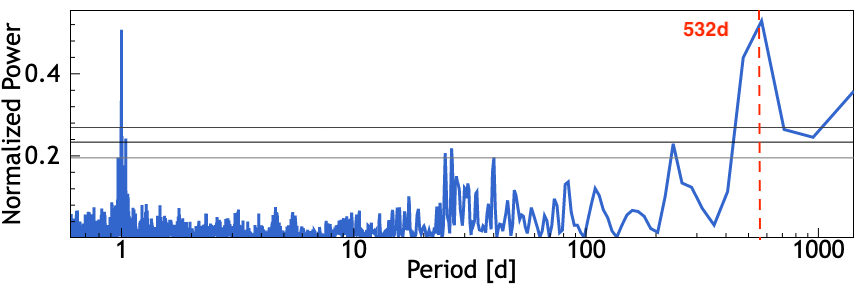}
    \caption{Sequential periodograms of SPIRou RV data of GJ~1148. From top to bottom: 41d peak and the 532d peak. The horizontal lines show the 10\% (bottom), 1\% (middle), and 0.1\% (top) FAP values. \label{fig:GJ1148spirouP}}
\end{figure}

\subsection{Multiplanet system hosted by GJ\,1148}
GJ\,1148 hosts a known two-planet system consisting of two Saturn-like eccentric planets with 41.4d and 532d orbital periods \citep{trifonov2018}. In addition to SPIRou data alone, we analyzed the full set of RV data from HIRES \citep{talor2019}, CARMENES \citep{ribas2023}, and SPIRou combined.

We applied the \texttt{wapiti} post-processing correction \citep{ouldelhkim2023} to the SPIRou residuals after removing the literature planetary signal of both planets \citep{trifonov2020}. As explained in section 2.3, this data-driven algorithm separates the principal components of the per-line RVs to remove seasonal systematics that plague specific lines (e.g., those most sensitive to tellurics). After correcting for and re-injecting the literature signal, the 104 SPIRou RVs of GJ~1148 have a standard deviation of 41 m/s. The known 41d period signal is detected with a $\log_{\rm FAP}$ $<$ -15.9 and the 532d signal is detected with a $\log_{\rm FAP}$ of -10.6 (or -7.0 when \texttt{wapiti} is not applied). 
The Lomb-Scargle periodograms of SPIRou RV data alone are shown in Figure \ref{fig:GJ1148spirouP}.

\begin{figure*}
    \centering
\includegraphics[width=0.68\hsize]{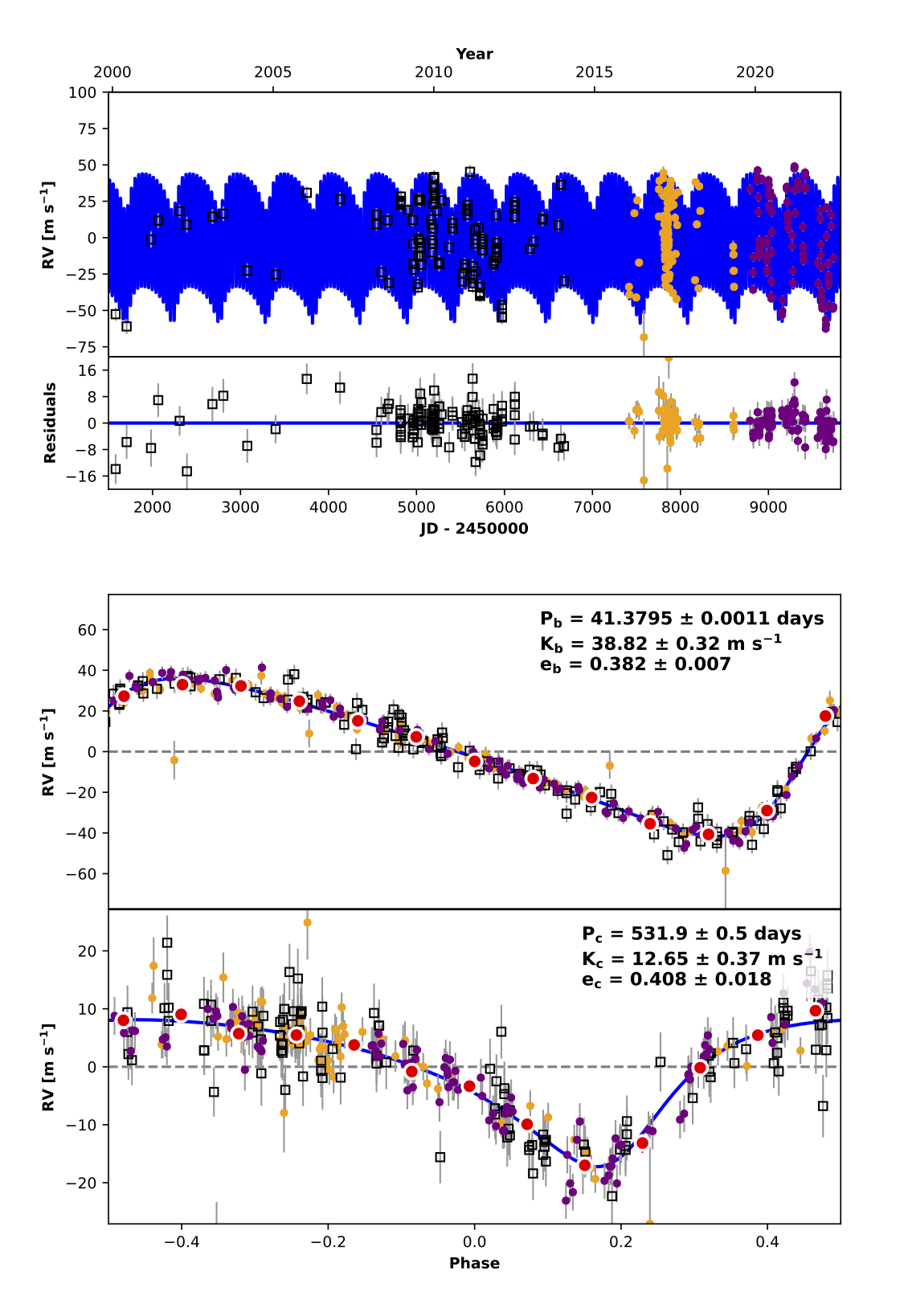}
\includegraphics[width=0.6\hsize]{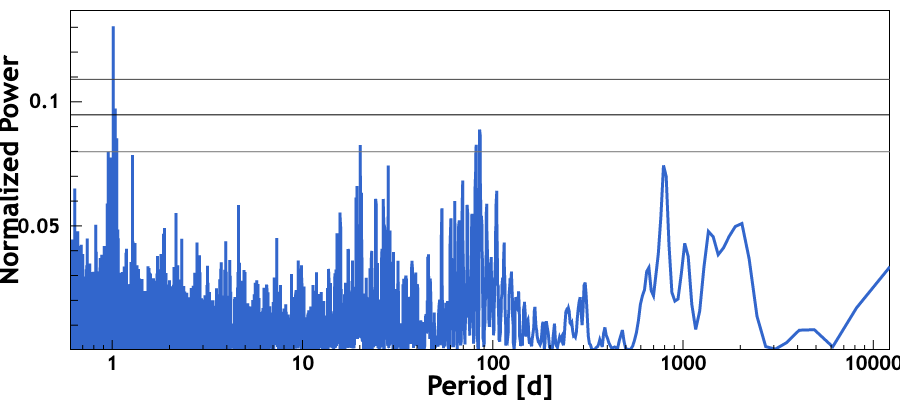}
    \caption{RVs of GJ~1148 and the two-planet fit obtained when using HIRES (open black squares), CARMENES (yellow dots), and SPIRou (purple dots) data. From top to bottom: Full signal as a function of time, and the signal phase folded at the period of planet b and planet c. The red dots show the binned data on the phase-folded plots. The \texttt{RadVel} two-planet model is shown as a blue line. The bottom plot shows the periodogram of the residuals, exhibiting an additional peak at the 85\,d period. The horizontal lines show the 10, 1, and 0.1\% FAP levels.}\label{fig:gj1148_2p}
\end{figure*}

When combined with CARMENES \citep{ribas2023} and HIRES \citep{talor2019} optical RV data, the two periods become stronger in the Lomb-Scargle periodogram  of the residuals of the two-planet fit, around periods of 20.1\,days ($log_{\rm FAP}$ of -0.81) and 85.5\,days ($log_{\rm FAP}$ of -1.914). We adjusted the whole data set (SPIRou, CARMENES, and HIRES) with two models: the two-planet model and a three-planet model including the most significant remaining candidate signal at the 85\,d period. We used \texttt{RadVel} \citep{fulton2018}  to optimize the fit, ran the Monte Carlo Markov chain (MCMC) analysis to explore the parameter domain (\texttt{RadVel} uses \texttt{emcce} from \citet{foreman2013}), and compared the models. Table \ref{tab:GJ1148params} compares the posterior parameters for the two-planet and three-planet models. The chosen prior for the period of the three signals is a Gaussian prior with a five-day width. The three-planet model is favored by the Bayesian model comparison, with an improvement of 17 in the logarithm of the marginal likelihood. However, at the moment, we reject the idea that this signal originates from a planet. This signal is not seen by individual instruments and only appears when three data sets are combined, which must trigger additional caution. 
In addition, because the period is close to the putative photometric period \citep{diezalonso2019}, this small signal might also be due to magnetic activity. An argument against this is that magnetic activity is not even seen in the magnetic field time series or appears significantly in other activity proxies (see next section). 
Finally, with the current best-fit parameters of the three-planet model, the orbits of the two inner planets would quickly be unstable. 
The other less significant candidate signals at a period of 20\,d are not confirmed by a model comparison analysis and were not considered further.
Figure \ref{fig:gj1148_2p} shows the SPIRou RV data in addition to the literature data from \citet{talor2019} and \citet{ribas2023}, together with the two-planet fit with \texttt{RadVel} and the periodogram of residuals. 

\begin{figure}
    \centering
\includegraphics[width=0.9\hsize]{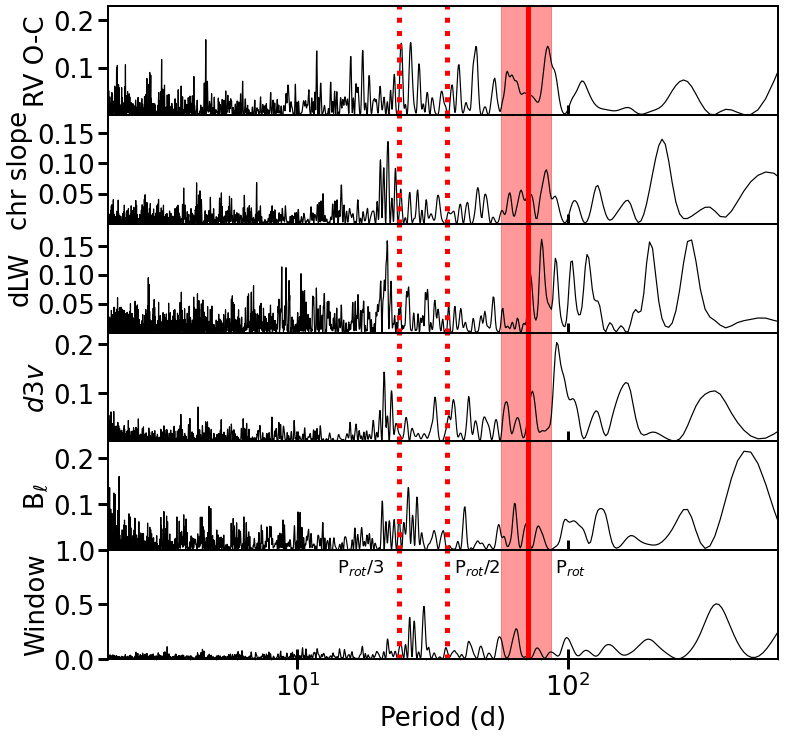}
    \caption{Periodograms of RV residuals (top row), SPIRou ancillary LBL data (next three rows), longitudinal magnetic field, and spectral window (bottom row) for GJ\,1148. The red zone shows the estimated rotation period \citep{diezalonso2019} and its 3$\sigma$ range; the first two harmonics of this rotation period are also shown as dotted lines.}\label{fig:gj1148p}
\end{figure}

\begin{table}
\caption{Fit and derived parameters for the system GJ~1148}
\begin{centering}
\begin{tabular}{lc}
\hline
Parameter & b+c \\
\hline
  $P_{b}$ (d) & $41.3795\pm 0.0011$   \\
  $T\rm{peri}_{b}$ & $2455477.92\pm 0.13$  \\
  $e_{b}$ & $0.3824\pm 0.007$\\
  $\omega_{b}$ (rad)& $-1.762\pm 0.022$  \\
  $K_{b}$ (m/s)& $38.82\pm 0.33$ \\
  $P_{c}$ (d) & $531.89\pm 0.5$ \\
  $T\rm{peri}_{c}$ & $2455966.5\pm 6.1$  \\
  $e_{c}$ & $0.408\pm 0.019$ \\
  $\omega_{c}$ (rad) & $-2.65\pm 0.068$ \\
  $K_{c}$ (m/s) & $12.65\pm 0.37$  \\
\hline
  $\gamma_{\rm SPIRou}$ (m/s) & $1.55\pm 0.34$ \\
  $\gamma_{\rm HIRES}$ (m/s) & $1.99\pm 0.43$ \\
  $\gamma_{\rm CARM}$ (m/s) & $-3.74\pm 0.35$   \\
  $\sigma_{\rm SPIRou}$ (m/s) & $2.7\pm 0.3$    \\
  $\sigma_{\rm HIRES}$ (m/s) & $3.8\pm 0.4$  \\
  $\sigma_{\rm CARM}$ (m/s) & $2.2\pm 0.3$  \\
\hline
  $m_{b}$ (M$_\oplus$)       &   $98.2\pm 2.9$    \\
  $a_{b}$ (au)               &    $0.17 \pm 0.02$     \\
  $m_{c}$ (M$_\oplus$)       &   $ 74.0\pm 2.9$          \\
  $a_{c}$ (au)               &    $ 0.914\pm0.013 $       \\
  $\ln{\mathcal{L}}$        &  -865.40         \\
\hline
\label{tab:GJ1148params}
\end{tabular}
\end{centering}
\end{table}

\subsection{Large-scale magnetic field of GJ~1148}

\begin{figure}
    \centering
\includegraphics[width=\hsize]{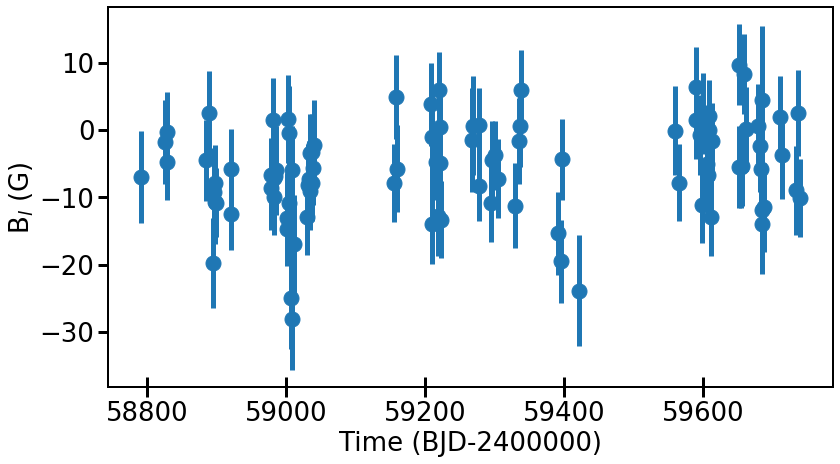}
     \caption{Measurements of the longitudinal magnetic field as a function of time, showing no periodic variation.}\label{fig:gj1148bl}
\end{figure}

Although the longitudinal magnetic field is detected in most visits of GJ~1148 \citep[see Figure C.10 in][]{fouque23}, no periodicity is found in the $B_\ell$ time series. The mean error on the $B_\ell$ values is 6.2 G, and the standard deviation of the time series is 7.0 G. There are too many unknown parameters (precise rotation period and inclination) for us to try reconstructing a meaningful magnetic topology from the circular polarization profiles, but the low-amplitude variability already suggests that the magnetic field is almost axisymmetric, or that the star is viewed almost pole-on (or both). Figure \ref{fig:gj1148bl} shows the $B_\ell$ variations with time. 
Although the main peak in the B$_\ell$ periodogram is at 431 days (Figure \ref{fig:gj1148p}), we do not interpret this variation as rotational modulation. The rotation period is estimated to be 71.5$\pm$5.1 days from a reanalysis of photometric HATNet data ( \citet{hartman2011}) and a 3\,mmag variation ( \citet{diezalonso2019}). The longer-term variation, however, is probably real because it is not present in the null calculated in spectropolarimetry. It may represent other long-term variability on the stellar surface.

The activity proxy dLW barely shows a peak close to the estimated rotation period, as shown in Figure \ref{fig:gj1148p}, whereas other proxies (chromatic slope, $d3v$, and longitudinal field) do not show a significant power at this period. The $d3v$ parameter may show a periodicity near 100d, outside the 3$\sigma$ interval of the previously estimated rotation period. The RV residuals after removal of the two-planet model do not exhibit an activity signal either (top row of this figure). The activity level of this star is low and/or stable, and the RMS on the RV jitter values is about 2 to 3 m/s for all instruments and wavelength ranges (Table \ref{tab:GJ1148params}). 

\section{Discussion and conclusion}
\label{sec:conclusion}

\begin{figure*}
    \centering
\includegraphics[width=\hsize]{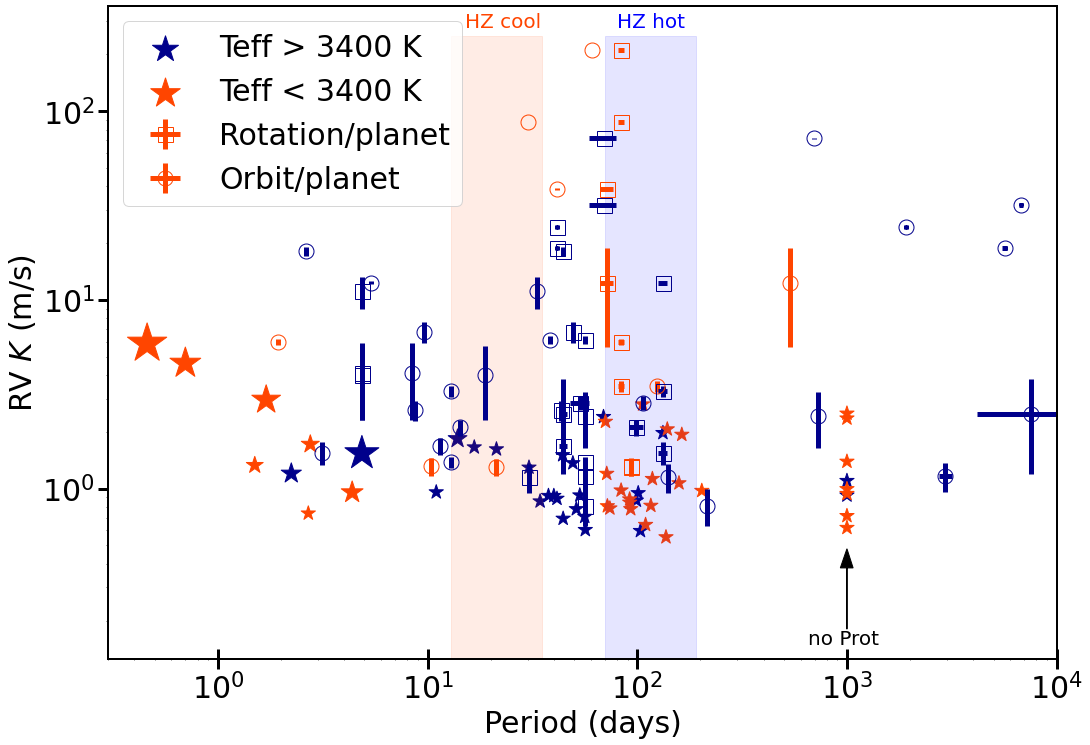}
\caption{Planet detectability from RV error measurements and number of visits as a function of the rotation period, compared to the habitable-zone period domains. Orange and blue depict the stellar host temperature below or above 3400~K. Stars without measured or literature values for the rotation period are arbitrarily placed at an abscissa value of 1000 days. Known planets in our sample are also included twice, first using their orbital period (open circles), and a second time using the rotation period of their star (open squares). }\label{fig:kprot}
\end{figure*}

We comprehensively described the blind planet-search program of the SPIRou legacy survey that was conducted from February 2019 to June 2022 at the CFHT. Collecting more than 7400 spectro-polarimetric visits on 57 stars, this survey is one of the most complete data collections of NIR RVs and has targeted nearby M stars of various spectral types. In this first analysis, we have gathered global characteristics of the survey and the data and a focused analysis of a few individual data sets. Sixteen known exoplanetary systems are included in the sample, and we focused here on SPIRou data acquired on two of them with large-amplitude signals: GJ~876, and GJ~1148.\\

The first conclusion is that achieving an RV accuracy of about 2.2 (resp. 4.0) m/s with SPIRou on a slowly rotating M dwarf requires an S/N of at least 100 per 2.28~\kms\ pixel in the H band for stars cooler (hotter) than 3400~K. The RV errors thus increase with the effective temperature of the star. When summing data over a very long sequence on one of the hottest stars in our sample (GJ~15A and 3610~K), a floor noise value of 0.7 m/s was found that may be due to instrumental limitations, for instance, detector noise or modal noise. We conducted comparisons with values adopted in \citet{reiners2020} for stars in our effective temperature range that were observed at values similar to our observed S/N. These calculations are reported in Fig. \ref{fig:phn} (bottom) with crosses. The expected photon-noise errors are two to three times better than observed. This shows that our way of estimating photon-noise RV uncertainties may be further improved, and/or that our current RV error does not reflect photon noise alone. Investigating this difference is beyond the scope of this article, but is certainly worth exploring further.\\

Then, using the four consecutive spectra obtained in a polarimetric sequence, we searched for specific RV patterns. These sequences combined two effects in SPIRou: first, the rhomb configuration changes from one exposure to the next; and second, the signal read by the detector seems to gradually stabilize during a four-exposure sequence due to persistence effects. Based on a variety of data from stellar and calibrations sequences, we conclude that the second effect with the signal stabilization largely dominates within a sequence. As discussed in the literature, persistence consists of incomplete resets between exposures, leaving out trapped charges that evaporate during the successive exposures \citep{baril2008,smith2008}. For RV studies, persistence depends not only on the illumination history of the detector, but also on the velocity differences from star to star, and it also depends on the spatial distribution of the defect over the detector array. It is therefore a very complex model to draw \citep{artigau2018}. Very subtle effects as seen at high flux in the four-exposure sequences, or greater ones on faint star spectra (not shown in this study), will help us to characterize the persistence in SPIRou RV data over time. 

The polarimeter has a mild influence on separate science channels, which compensates for and self-cancels in the summed science spectrum where RVs are calculated. However, there is a notable RV offset between the first and last exposure of a sequence, notably visible on the Fabry-P\'erot spectra in the reference channel at S/N values lower than 100. As a variable density is used to adjust the reference channel flux to the star magnitude, this offset is constant for a given star and should not introduce a systematic jitter. This gradual offset thus only complicates the use of FP nightly measurements to completely characterize the instrumental drift. The smaller offset seen in the stellar spectra of about -1 m/s for all stars, even those observed at very high S/N, may be due to the same physical effect complicated by different RV contents and systemic velocities from one star to the next. Trying to correct for a systematic effect at this level is undoubtedly very difficult. Further effort will be devoted to understanding, characterizing, and correcting for this small effect in the future. In the meantime, we recommend to combine all four RVs of stars and FP and correct for the drift, star by star, as was done for the present analysis.\\

Measuring the stellar RVs in small chunks of the spectra allows us to see the variation in RV precision as a function of wavelength in the wide SPIRou domain. RVs in the $H$ and $K$ bands, where the throughput is also the highest, contain most RV information in SPIRou, with relative weights twice higher than in the bluer bands. The $K$ band has a slightly lower accuracy for the hottest stars in our sample, above 3700~K. In turn, the $Y$ band in SPIRou achieves a better accuracy for stars cooler than 3200~K or hotter than 3500~K.\\

To conclude, we calculate the potential of the SPIRou planet-search survey by combining the RV error measurement per star and the number of collected measures, including a white-noise term of 2 m/s. In Figure \ref{fig:kprot} we show an estimate of the lower RV amplitude that SPIRou could detect with an S/N of 5 as a function of the rotation period of the star \citep{fouque23} (stars). It assumes that the RV final dispersion has a similar amplitude as the mean error, when all large signals are removed (known planets and rotationally modulated signal). This needs yet to be proven. The sample is split into the coolest part and the hottest part, with a limit of 3400~K. The rough location of the habitable zones \citep{Kopparapu2013} for these two types of stellar host temperatures is also shown for comparison. It is fortunate that in this specific sample, the rotation periods of stars do not usually coincide with periods corresponding to their habitable zone. For instance, the rotation period of most stars hotter than 3400~K is lower than the period range corresponding to their HZ. The rotation periods of our coolest stars are either much shorter or much longer than the period range of their HZ. This coincidence should help us distinguish potential stellar signals from planetary signals in the HZ of their hosts if any are detected. 
The known planets (systems listed in Table \ref{tab:knownplanets}) are also shown in this plot in two ways: with the abscissa put to the rotation period of their host (open squares), or to their orbital period (open circles). In both ways, the symbols are colored with the effective temperature of their host star. this mainly shows that the detectability zone of SLS-PS data corresponds to a lower envelope with respect to known planetary signals with the hypotheses above. A thorough investigation of all RV time series will be the subject of future work.

The large number of ongoing surveys targeting nearby M dwarfs offers an important overlap for future studies. Combining optical and NIR data may not be an easy task for moderately active stars, but it is feasible for quiet stars, as shown in a few examples in this paper, providing that the RV precision is sufficient. In forthcoming work, other examples of a combined data analysis will further explore this avenue.\\

Three individual data sets were then presented. To start, an engineering five-hour-long sequence in 48 consecutive polarimetric sequences on GJ~15A was analyzed. These very long sequences are extremely rarely performed at the telescope or contain variable planetary effects. This sequence, free of such signals, was also used to test the atmospheric transmission pipeline ATMOSPHERIX \citep{debras2023}. In the RV context, this long series has mostly shown a very stable star and instrument, with a dispersion of 1.42\,m/s at the level of photon noise precision. A single bump in the stellar line width is observed that lasts less than one hour. It is unclear whether this variation is intrinsic to the star (as no other parameter shows a similar pattern, except the S/N) or instrument/pipeline related (e.g., due to near-field effects). During this sequence, the RV in the $Y$ band varies slightly more than RVs in other bands, although within the precision. The line-of-sight magnetic field is also stable during this sequence, which allowed us to combine all 48 polarimetric spectra and reach a sensitivity at a level of 2.2 10$^{-5}$ (see section 4.2).\\

The second SPIRou data set that was analyzed in this paper is the set of 88 spectropolarimetric visits of the mid-M dwarf GJ\,876, which is host to a resonant four-planet system. One objective was to update the dynamical analysis of this system using the 24-year time span while combining HIRES, HARPS, APF, PFS, CARMENES, and SPIRou data. This updated dynamical analysis further confirmed that the three outer planets are deeply embedded in the Laplace mean-motion resonance and give results that are consistent with the previous analysis. The eccentricity of planet d is yet slightly lower than previous estimates (1.4$\sigma$ lower than the value derived by \citet{trifonov2018}), which agrees with the trend of favoring a shorter circularization timescale and lower tidal quality factor, as noted in \citet{puranam2018}. When a lower stellar mass derived from SPIRou data and the global RV analysis are used, the mass of the planets also mostly changed compared to literature values: while the masses of planets b, c, and d (the three inner planets) are now lower by a factor 0.88, the mass of the outer planet e is similar to the literature because a larger semi-amplitude was derived in our study. Finally, our dynamical analysis allowed us to derive a well-constrained Laplace resonance angle of 30.6$_{-5.8}^{+4.6}$ degrees. This value is in between the values of \citet{Nelson2016} and \citet{Millholland2018} and to 1$\sigma$ compatible with both.

The second objective was to derive the magnetic map through Zeeman-Doppler imaging, which was performed for both observational seasons separately due to a notable evolution of the field over the time span of observations. The large-scale topology of the GJ~876 magnetic field is characterized by a 30~G dipole. This topology is typical for this type of star; as a comparison, Proxima Cen also has a tilted dipole field, but its amplitude is larger \citep{klein2021a}. The dipole of GJ~876 is tilted with respect to the rotation axis of the star, and it is interesting to note that the tilt has significantly evolved on the timescale of one year, from 30 to 60$^\circ$. Similar behaviors are seen in other M stars, such as the active star AD Leo \citep{bellotti2023} or a few more moderately active stars such as GJ\,1289 (Lehmann et al, in prep). A global analysis of M-star magnetic topologies and their evolution will be performed from SLS observations and will allow a better understanding of the magnetic environment, stellar wind properties of M stars, and star-planet interactions, as can be theoretically described \citep[e.g.,][]{vidotto2014,strugarek2015,garraffo2022}. In this respect, the known inner planet (d) of GJ\,876, orbiting at a 1.94~d period, strongly interacts with the close-in star and has a local magnetic field of about 10\,mG. This is a similar magnitude to other close-in planets such as the hot Jupiter HD~189733 b \citep{fares2017} or the super-Earth Kepler-78 b \citep{moutou2016,strugarek2019}.  \\

Last, GJ\,1148, another star hosting a multiple-planet system, was analyzed. In this system, SPIRou data easily detected the two giant planets in 41\,d and 532\,d orbital periods. Combining the analysis with CARMENES and HIRES data, we were able to refine the outer planet period and semi-amplitude by a factor of $\sim$2 on their accuracy. In addition, a third signal may be detected when all three data sets are combined. This 3$\sigma$ detection is not considered a robust planet signal. The planet scenario is not likely due to stability arguments: A system close to resonance with such massive and eccentric planets is highly unstable. It is even very likely that planet-planet scattering has led to the current configuration of the system and cleared out other planets, as discussed in \citet{trifonov2020}. 
For this stellar host, although individual polarimetric visits show the magnetic field detection with a mean value of $-5.7\pm6$\,Gauss, no rotational modulation is observed. This is not a favorable configuration to derive the magnetic topology of the star because the problem is highly degenerate.\\

Future work on the Planet-Search SPIRou data set includes the analysis of all times series, in search for the RV, spectral, and spectropolarimetric properties. Making progress in the data analysis also gives us more confidence in the data reduction and will allow us to make the full data set public in the near future for its legacy value.

\begin{acknowledgements}
This project received funding from the European Research Council under the H2020 research \& innovation program (grant 740651 NewWorlds).
\par
E.M. acknowledges funding from FAPEMIG under project number APQ-02493-22 and research productivity grant number 309829/2022-4 awarded by the CNPq, Brazil.
\par
We acknowledge funding from Agence Nationale pour la Recherche (ANR, project ANR-18-CE31-0019 SPlaSH).
\par
X.D. and A.C. acknowledge funding by the French National Research Agency in the framework of the Investissements d’Avenir program (ANR-15-IDEX-02), through the funding of the “Origin of Life” project of the Universit\'e Grenoble Alpes.
\par
We thank the Swiss National Science Foundation (SNSF) and the Geneva University for their continuous support to our planet search programs. This work has been in particular carried out in the frame of the National Centre for Competence in Research PlanetS supported by the SNSF. This publication makes use of The Data \& Analysis Center for Exoplanets (DACE), which is a facility based at the University of Geneva dedicated to extrasolar planets data visualisation, exchange and analysis. 

\end{acknowledgements}

\bibliographystyle{aa}
\bibliography{00_references}

\begin{appendix}
\section{Observation log}
Table \ref{tab:slsobs} lists the observation parameters and some stellar parameters.

\begin{table*}[]
    \centering\small
    \caption{List of stars observed in the framework of the SPIRou legacy survey planet-search program}
    \begin{tabular}{lcccccccllc}
\hline 
Star         & Mass$^a$ & M/H$^a$ & distance & T$_{\rm eff}$$^a$ & mH & $v \sin i$$^b$ &   Nvis$^c$  & Texp$^d$ &   S/N$^d$  & $\sigma_{rv}$$^d$   \\
             & M$_\odot$ &  & (pc) & (K)      &     & (km/s)&           &  (s) & per 2.28 km/s pixel   &(m/s) \\\hline
GJ1002       &0.12      &-0.33  &4.85  &2980 & 7.792 & $<$2.0  &  141 &   246 (150 -- 301) &       99 (70 -- 125) & 2.86\\
GJ~15A       &0.39      &-0.33  &3.56  &3611 & 4.476 & $<$2.0  &       238 &    61 (fixed)        &      268 (177 -- 317) & 3.30\\
GJ~15B       &0.16      &-0.42  &3.56  &3272 & 6.191 & $<$2.0 &       180 &    68 (50 -- 72)   &       99 (70 -- 122) & 3.81\\
GJ~1012      &0.35      &0.07   &13.38 &3363 & 7.504 & $<$2.0  &       139 &   193.2 (128 -- 228) &      98 (70 -- 117) & 3.62\\
GJ~48        &0.46      &0.08   &8.23  &3529 & 5.699 & $<$2.0  &       194 &    61 (56 -- 61)   &      125 (70 -- 152) & 3.57\\
GJ~169.1A    &0.28      &0.13   &5.52  &3307 & 6.012 & $<$2.0  &       173 &    61  (fixed)       &      103 (70 -- 126) & 3.94\\
GJ~205       &0.58      &0.43   &5.70  &3771 & 4.149& $<$2.0  &       155 &    61 (50 -- 61)   &      287 (177 -- 347) & 3.43\\
GJ~3378      &0.26      &-0.05  &7.73  &3326 & 6.949& $<$2.0  &       173 &   136 (89 -- 150)  &      106 (70 -- 133) & 3.67\\
GJ~251       &0.35      &-0.01  &5.58  &3420 & 5.526& $<$2.0  &       157 &    61   (fixed)      &      139 (106 -- 166) & 3.56\\
GJ~1103      &0.19      &-0.03  &9.27  &3170 & 7.939& $<$2.0  &        65 &   253 (201 -- 312) &       95 (70 -- 113) & 3.32\\
GJ~1105      &0.27      &-0.04  &8.85  &3324 & 7.133& $<$2.0  &       162 &   160 (111 -- 173) &      105 (70 -- 130) & 3.88\\
GJ~1111  (DX Cnc)&0.10 &-0.15  &3.58  &2997 & 7.617& 10.5 &        13 &   223 (fixed)        &       70 (60 -- 82) & 9.18\\
PM~J08402$+$3127 &0.28&-0.08 &11.23  &3347 & 7.561& $<$2.0  &     136 &   204 (162 -- 245)  &      96 (70 -- 110) & 3.89\\
GJ~317       &  0.42&0.23       &15.18  &3421 & 7.321& $<$2.0  &        68 &   188 (139 -- 262) &      108 (85 -- 129) & 4.04\\
GJ~338B      &0.58      &-0.08  &6.33  &3952 & 4.043& $<$2.0  &        54 &    61 (fixed)        &      234 (155 -- 297) & 4.59\\
PM~J09553-2715 &0.29&-0.03&10.90 &3366 & 7.433& $<$2.0  &         71 &   210 (150 -- 301)  &     110 (85 -- 141) & 3.86\\
GJ~382       &0.51      &0.15   &7.71  &3644 & 5.26& $<$2.0  &       105 &   69 (50 -- 95)     &     155 (109 -- 396) & 4.17\\
GJ~388 (AD Leo) &0.42   &0.24&4.97  &3449 & 4.843& 3.0 &        74 &    61 (fixed)        &      172 (123 -- 221) & 3.69\\
GJ~3622      &0.10       &-0.41 &4.56  &3031 & 8.263& 2.1 &        83 &   372 (301 -- 451) &       92 (63 -- 109) & 2.81\\
GJ~406       &  0.11&0.17&2.41  &2898 & 6.482& $<$2.0  &       165 &   111 (67 -- 128)  &      114 (77 -- 148) & 2.62\\
GJ~408       &0.38      &-0.09&6.75  &3487 & 5.76& $<$2.0  &       169 &    78 (56 -- 89)   &      135 (89 -- 172) & 3.68\\
GJ~410       &0.55      &0.05&11.94  &3818 & 5.899& 2.6 &       129 &    72 (56 -- 84)   &      127 (84 -- 151) & 5.77\\
GJ~411       &0.39      &-0.38&2.55  &3589 & 3.64& $<$2.0  &       180 &    61 (fixed)        &      355 (212 -- 441) & 2.55\\
GJ~412A      &0.39      &-0.42&4.90  &3620 & 5.002& $<$2.0  &       176 &    61 (fixed)        &      177 (108 -- 221) & 3.72\\
GJ~1148      &0.34      &0.11&11.03       &3354 & 7.069& $<$2.0  &       104 &   139 (100 -- 150) &       96 (70 -- 111) & 3.68\\
GJ~436       &0.42      &0.03&9.77  &3508 & 6.319& $<$2.0  &        75 &   179 (111 -- 245)  &     151 (106 -- 202) & 3.57\\
GJ~445       &0.24      &-0.24&5.25  &3356 & 6.217& $<$2.0  &        94 &    83 (61 -- 95)   &      112 (71 -- 140) & 3.30\\
GJ~447       &0.18      &-0.13&3.38  &3198 & 5.945& $<$2.0  &        55 &    89     &     129 (76 -- 166) & 3.22\\
GJ~1151      &0.17      &-0.16&8.04  &3178 & 7.952& $<$2.0  &       153 &   275 (195 -- 301) &       97 (70 -- 117) & 3.61\\
GJ~1154      &0.18      & - &8.09  &3078 & 7.86& 6.1 &        32 &   400 (396 -- 401) &      129 (88 -- 154) & 5.30\\
GJ~480       &0.45      &0.26&14.26  &3509 & 6.939& $<$2.0  &       108 &   149 (106 -- 173)  &     108 (78 -- 124) & 4.15\\
GJ~514       &0.50      &-0.07  &7.62  &3699 & 5.3& $<$2.0  &       153 &  67 (50 -- 95)      &     157 (105 -- 239) & 4.40\\
GJ~536       &0.52      &-0.08  &10.42  &3800 & 5.93& $<$2.0  &        12 &   121 (106 -- 123)  &     148 (125 -- 159) & 3.73\\
GJ~581       &0.31      &-0.07  &6.30 &3406     & 6.095& $<$2.0  &         26 &   122 (89 -- 128)   &     123 (86 -- 152) & 3.92\\
GJ~617B      &0.45      &0.20   &10.76  &3525 & 6.3& $<$2.0  &       146 &   100 (72 -- 117)   &     119 (85 -- 152) & 4.14\\
GJ~687       &0.39      &0.01   &4.55  &3475 & 4.766& $<$2.0  &       205 &    61 (fixed)         &     199 (114 -- 243) & 2.81\\
GJ~699       &0.16      &-0.37  &1.83  &3311 & 4.834& $<$2.0  &       245 &    61 (fixed)         &     197 (112 -- 242) & 2.53\\
GJ~4063      &0.35      &0.42   &10.89  &3419 & 6.53& $<$2.0  &       213 &    89 (61 -- 100)   &      99 (70 -- 122) & 4.51\\
GJ~725A      &0.33      &-0.26  &3.52  &3470 & 4.741& $<$2.0  &       212 &    61 (fixed)         &     210 (125 -- 255) & 2.81\\
GJ~725B      &0.25      &-0.28  &3.52  &3379 & 5.197& $<$2.0  &       208 &    61 (fixed)         &     158 (86 -- 195) & 2.95\\
PM~J18482$+$0741 &0.14&-0.02    &7.62 &3080     & 8.261& 2.4 &       98 &   227 (162 -- 234)  &      76 (56 -- 111) & 5.17\\
GJ~752A      &0.47      &0.11   &5.92  &3558 & 4.929& $<$2.0  &       128 &    60 (50 -- 61)    &     175 (93 -- 230) & 3.59\\
GJ~1245B     &0.12      &-0.05 &        4.66 &2944      & 7.728 & 7.0 &        18 &   237 (206 -- 245)  &      76 (60 -- 87) & 7.35\\
GJ~1256      &0.29      & 0.02  &9.53  &3270 & 8.075& $<$2.0  &        15 &   376 (306 -- 396)  &      85 (69 -- 96) & 3.85\\
PM~J21463$+$3813 &0.18  &-0.38  &7.05  &3305 & 7.465& $<$2.0  &     180 &   195 (139 -- 228)  &      99 (71 -- 122) & 3.24\\
TYC~3980-1081-1      & 0.48 & -  &8.12  & 3262  & 5.86& $<$2.0  & 69 &    60 (56 -- 61)    &     106 (71 -- 137) & 7.40\\
GJ~846       &0.57      &0.07   &10.57  &3833 &5.562& $<$2.0  &       195 &    83 (56 -- 123)   &     157 (108 -- 231) & 4.92\\
GJ~849       &0.46      &0.35   &8.81  &3502 & 5.899    & $<$2.0  &       196 &    61 (fixed)         &     114 (70 -- 142) & 4.40\\
GJ~4274      &0.20      &-0.07  &7.23  &3230 & 7.638& $<$2.0  &       18  &  214 (178 -- 245)   &      98  (75 -- 102) & 3.75\\
GJ~873 (EV Lac) &0.32   &0.04&5.05&3341 & 5.554& 3.5 &       171 &    61 (fixed)       &      133  (70 -- 165) & 4.35\\
GJ~876       &0.33      &0.15&4.67 &3366        & 5.349& $<$2.0  &        91 &   72 (50 -- 123)    &     158 (107 -- 225) & 3.16\\
GJ~880       &0.55      &0.26&6.87  &3702 & 4.8& $<$2.0  &       164 &    61 (fixed)         &     204 (126 -- 250) & 3.87\\
GJ~4333      &0.37      &0.25&10.59  &3362 & 6.771& $<$2.0  &       187 &   111 (72 -- 123)   &      97 (71 -- 123) & 3.81\\
GJ~905       &0.15      &0.05&3.16  &3069 & 6.247& $<$2.0  &       196 &    88 (61 -- 95)    &     115 (84 -- 137) & 2.83\\
GJ~1289      &0.21      &0.05&8.36  &3238 & 7.446& $<$2.0  &       207 &   198 (145 -- 217)  &      96 (71 -- 110) & 3.64\\
GJ~803 (AU Mic) &0.60&0.12      &9.72  &3660 & 4.83& 8.5 &       167 &  187 (123 -- 201)   &     373 (176 -- 469) & 7.70\\
GJ~1286      &0.12      &-0.23  &7.18  &2961 & 8.505& $<$2.0  &        97 &  436 (228 -- 479)   &      96 (70 -- 122) & 3.21\\
\hline
\end{tabular}
    \label{tab:slsobs}
\tablefoot{    
\tablefoottext{a}{Mass, metallicity, and effective temperature mainly come from \citet{Cristofari2022, cristofari2023}.}
\tablefoottext{b}{Projected rotational velocity values come from \citet{fouque2018}.}
\tablefoottext{c}{The number of visits corresponds to nightly bins. A typical visit consists of four subexposures.}
\tablefoottext{d}{Exposure times, S/N (ranges in parenthesis), and RV photon-noise uncertainty $\sigma_{rv}$, are given per subexposure.}}
\end{table*}

\section{Dynamic analysis of GJ~876}
\label{app:GJ876}
\subsection{Details of the method}
We integrated the system using the Wisdom-Holman integrator \WHFast{} \citep{Rein2015a} from the \rebound{} package \citep{Rein2012a}.
The code units were days, AU, and $\mathrm{M}_\odot$.
We applied a symplectic corrector of order 3 and chose at a time step of $\Delta t=P_\mathrm{d}/15=0.13~$days.
These parameters limit the error in the RV modeling to 0.1 m/s while keeping the computing cost low.

We ran a Monte Carlo Markov chain (MCMC) using the package \texttt{emcee} \citep{Foreman-Mackey2013}.
We used the same differential evolution MCMC moves \citep{TerBraak2006} as \citetalias{Millholland2018}.
The parameters involved in the fit are the planet periods, RV semi-amplitudes, mean longitudes, eccentricities, and longitude of periapses, as well as a constant RV offset and jitter per data set.
We transformed the semi-amplitudes, mean longitudes, eccentricities, and longitudes of periapsis into the variables  $\sqrt{K}\cos\lambda$, $\sqrt{K}\sin\lambda$, $\sqrt{e}\cos\varpi$, and $\sqrt{e}\sin\varpi$ to avoid correlations.
In total, our model had 37 free parameters (five per planet, the system inclination and an offset, and a jitter per observational set), and we note the vector of parameters $\phi$.

Let $\vec{r}$ be the vector of the RV residuals after subtracting the $N$-body model and the instrument offsets with components
\begin{equation}
    r_i= \mathrm{RV_{obs}}(t_i) - \gamma_i -\mathrm{RV_{\emph{N}-body}}_i(t_i).
\end{equation}
Our log-likelihood is written as
\begin{equation}
    \ln \mathcal{L}(\vec{r}|\phi) = -\frac{1}{2}\sum_{i}\left( \frac{r_i^2}{\sigma_i^2}+\ln(\sigma_i^2)\right)
,\end{equation}
where 
$\sigma_i^2 = \sigma_{\mathrm{obs}}(t_i)^2+\sigma^2_{\mathrm{jit,inst}}$ is the sum of the uncertainty on the measure at epoch $t_i$ and a common jitter term of the instrument used.

We initialized the MCMC by drawing random states from the posterior distribution provided by \citetalias{Millholland2018}. 
Since the stellar mass value is different from the value used by \citetalias{Millholland2018}, we scaled the initial inclinations such that $M_{\mathrm{new}}^{1/3}\sin i_{\mathrm{new}} = M_{\mathrm{old}}^{1/3}\sin i_{\mathrm{old}}$.
For constant RV semi-amplitudes and periods, this scaling preserves the planet-to-star mass ratio that is the physical variable to which the dynamical RV fit is sensitive.
We used uniform priors and only enforced that the eccentricities remained lower than 1.

\subsection{Dynamical considerations}

\begin{figure}
    \includegraphics[width=\linewidth]{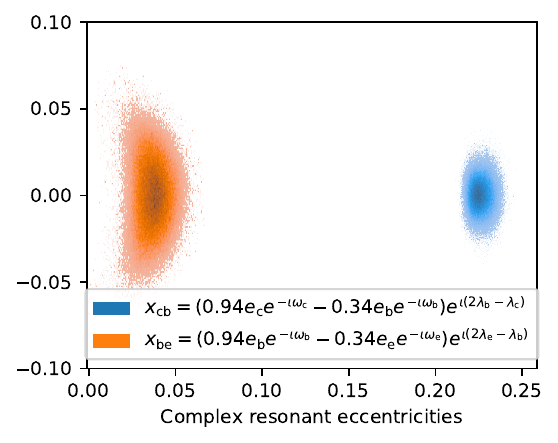}
    \caption{Probability density function of the resonant variables $x_\mathrm{cb}$ and $x_\mathrm{be}$ for 200 initial conditions drawn from the posterior distribution and integrated for $10^7$ days.}
    \label{fig:resangle}
\end{figure}

In order to study the dynamical behavior of the system, we integrated over $10^7$ days, 200 initial conditions drawn at random from the posterior distribution.
With this longer integration, we aimed in particular to study the behavior of the resonant angles between planets b, c, and e.
The existence of a resonance between two planets is traditionally determined by the libration of one or several resonant angles.
In the context of the 2:1 MMR, the resonant angles take the form
\begin{equation}
    \phi_{12,k} = 2\lambda_2-\lambda_1-\omega_k
,\end{equation}
where $\lambda_k$ is planet $k$ mean longitude.
In the case of longer resonant chains such as GJ876, the three-body Laplace angle
\begin{equation}
    \phi_\mathrm{Lap,cbe} = \lambda_\mathrm{c}-3\lambda_\mathrm{b}+2\lambda_\mathrm{e}
\end{equation}
also librates, which indicates that the system is trapped deep in the resonance \citepalias{Millholland2018}.
However, the resonant dynamics close to the resonance fixed point are dominated by a single linear combination of the eccentricities and resonant angles \citep{Hadden2019}.
As a result, the information given by the resonant angle alone can give a misleading representation of the actual dynamical state of the system.
Following \cite{Petit2020}, for the 2:1 MMR, the resonant variable for the pair of planet $(j,k)$ is
\begin{equation}
    x_{jk} = 0.94 e_1e^{\iota\phi_{jk,j}}-0.34e_2e^{\iota\phi{jk,k}}.
\end{equation}
When a system is in a resonant state, the variable $x_{jk}$ is confined close to the positive $x$ -axis in the vicinity of the center of libration \citep[\emph{e.g.}][]{Petit2020}.
We plot on Figure~\ref{fig:resangle} the probability density function of the resonant variables $x_\mathrm{cb}$ and $x_\mathrm{be}$ for the 200 initial conditions along their integration.
Both resonant variables are confined close to their respective libration center.
However, the trajectories still fill the region close to the libration center, which indicates that the dynamics are still chaotic, even though the system is in resonance.

We quantified the libration amplitude of the different resonant angles by computing the root mean square value along individual integrations using the same technique as \citetalias{Millholland2018}.
The results are given in Table~\ref{tab:resangle}.
Our results are within the confidence interval of the amplitudes found by \citetalias{Millholland2018}.
In particular, we note that the large libration amplitude of the angle $\phi_\mathrm{be,e}$ that is due to the low eccentricity of planet e that is weakly constrained by the dynamical fit has a negligible impact on the resonant variable $x_\mathrm{be}$.
We also confirm the very good constraint on the Laplace angle obtained by \citetalias{Millholland2018}.

\begin{table}
    \caption{Amplitude of the different resonant angles between planets c, b, and e.}\label{tab:resangle}
\begin{center} \begin{tabular}{ l c c}  \hline\hline Angle & Libration center (${}^\circ$) & Amplitude (${}^\circ$) \\ \hline
$\phi_\mathrm{cb,c}$ &  0 &${2.44}^{+0.49}_{-0.35}$\\ 
$\phi_\mathrm{cb,b}$ &  0 &${12.5}^{+1.7}_{-2.4}$\\ 
$\mathrm{arg}(x_\mathrm{cb})$ &  0 &${3.03}^{+0.54}_{-0.48}$\\ \hline
$\phi_\mathrm{be,b}$ &  0 &${28.8}^{+4.4}_{-5.3}$\\ 
$\phi_\mathrm{be,e}$ &  180 &${92.9}^{+6.}_{-9.4}$\\ 
$\mathrm{arg}(x_\mathrm{be})$ &  0 &${37.8}^{+4.7}_{-6.}$\\ \hline
$\phi_\mathrm{Lap,cbe}$ &  0 &${30.6}^{+4.6}_{-5.8}$\\ \hline
\end{tabular}\end{center}
\end{table}

\section{Impact of seeing on the stellar line width}
\begin{figure}
    \includegraphics[width=\linewidth]{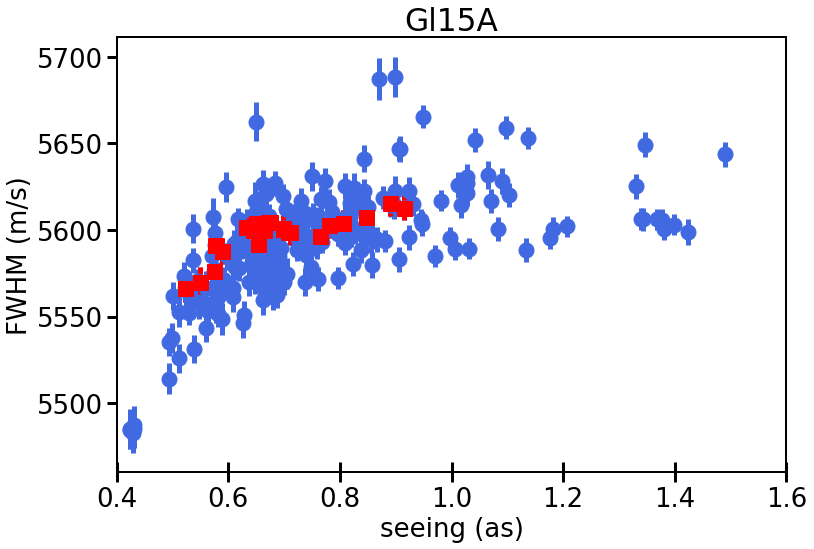}
    \includegraphics[width=\linewidth]{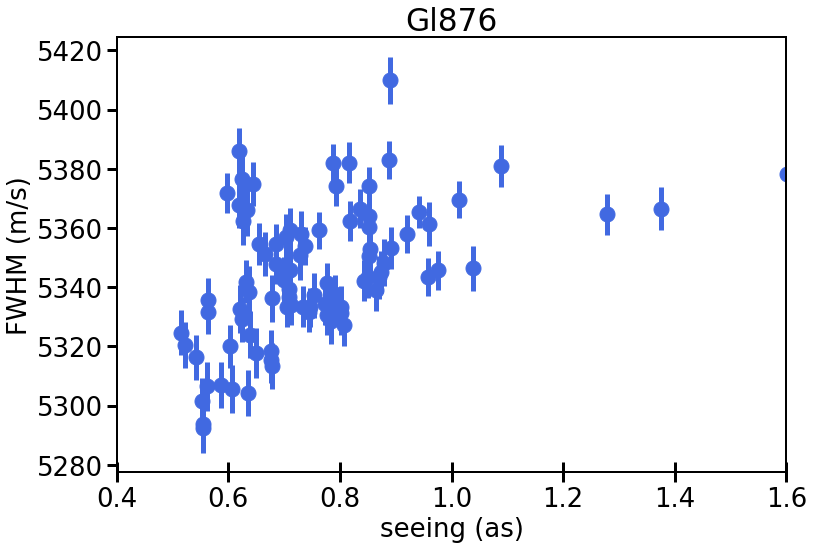}
    \includegraphics[width=\linewidth]{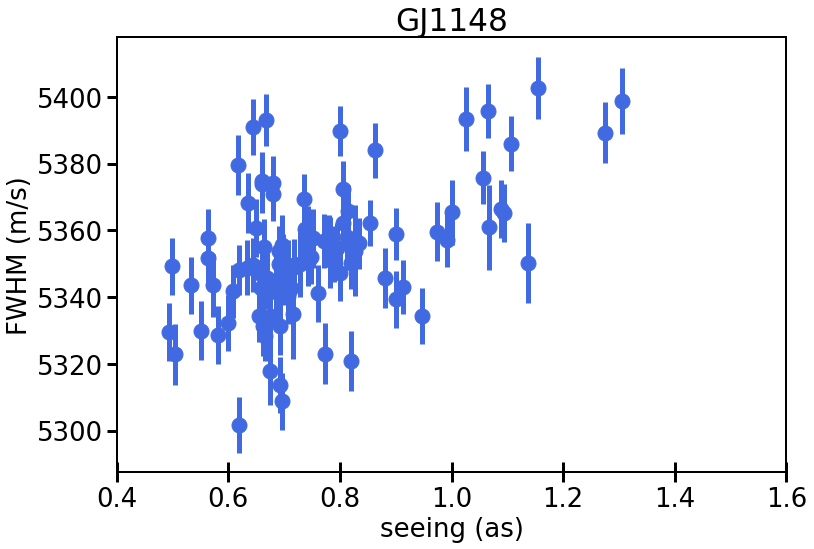}
    \caption{Stellar FWHM as a function of seeing for all three data sets. For GJ~15A, the red squares represent binned data from the five-hour sequence night.}
    \label{fig:seeing}
\end{figure}

\begin{figure}
    \includegraphics[width=\linewidth]{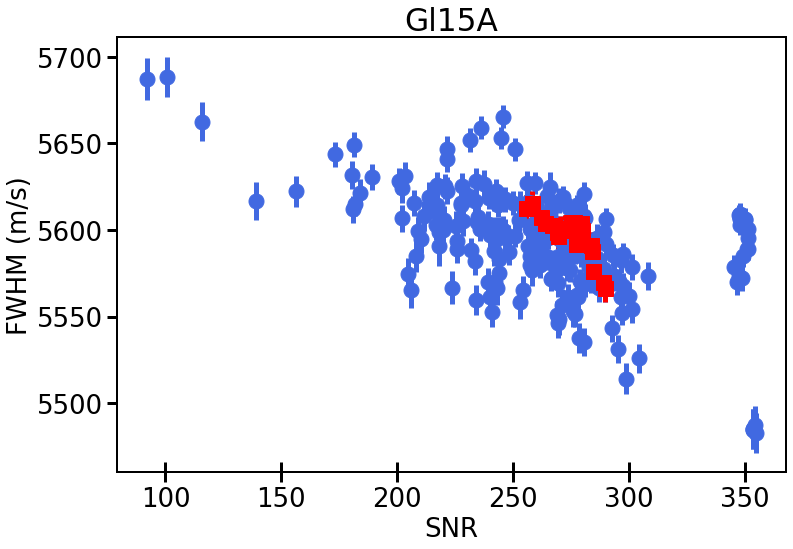}
    \includegraphics[width=\linewidth]{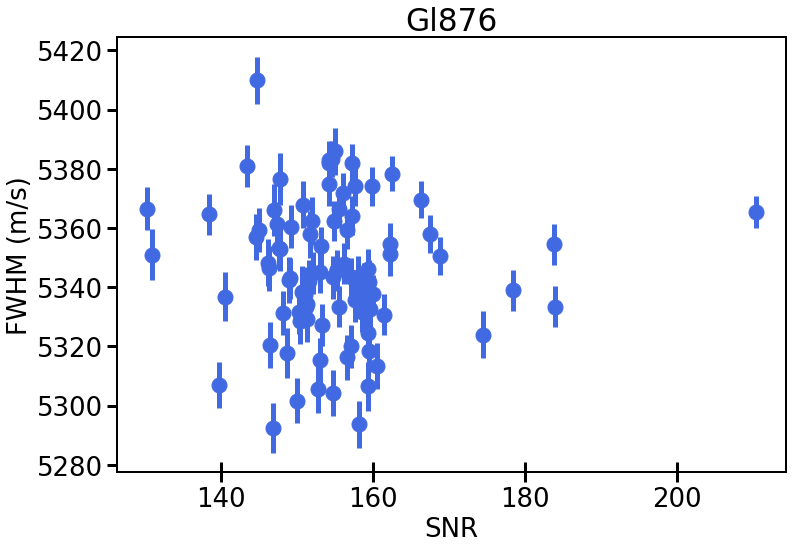}
    \includegraphics[width=\linewidth]{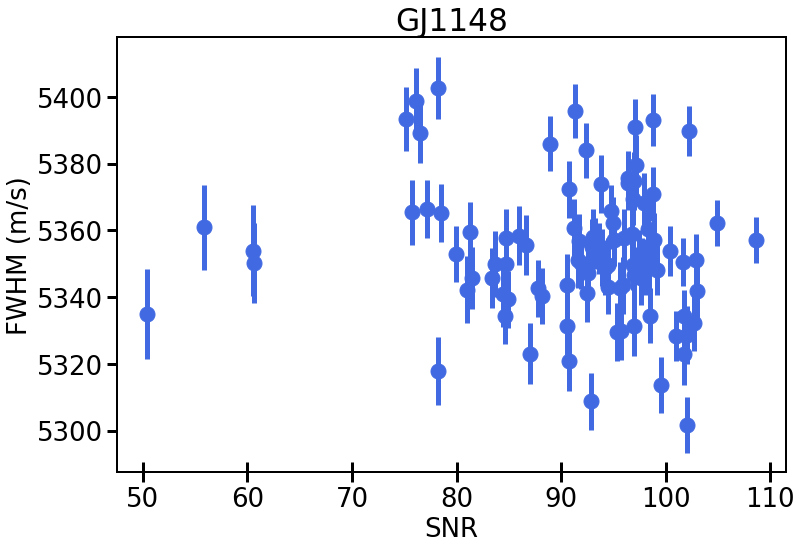}
    \caption{Stellar FWHM as a function of S/N for all three data sets. For GJ~15A, the red squares represent binned data from the five-hour sequence night.}
    \label{fig:snr}
\end{figure}

As observed in the five-hour sequence on GJ~15A, the mean differential line width varies slightly  with image quality and S/N, and these variations are correlated. Here we document the relation between these quantities further using data of GJ~15A, GJ~1148, and GJ~876 used in this article. The seeing value was estimated from the guiding images. The S/N is given per 2.28 km/s pixel in the $H$ band.
We draw the following conclusions: firstly, the stellar width variation is a few percents of the line width. Although this is significant, it remains an extremely faint effect. Secondly, the effect is strongest for GJ~15A (3.6\%), which is always observed at a fixed exposure time (and was observed at more extreme seeing values), while GJ~876 and GJ~1148 are observed in short but variable exposure times: they vary in relation with external conditions (e.g., seeing). The effect is weakest on GJ~1148, where the S/N range is also the lowest of the three stars. It is suspected that the seeing effect is directly related to an S/N effect on the differential line width. Figures \ref{fig:seeing} and \ref{fig:snr} show the mean FWHM of stellar lines for all visits as a function of seeing and S/N. While the seeing effect seems more clearly related to an S/N effect for GJ~15A, it is less clear for the other two stars. For a physical explanation and a clear separation between the seeing and S/N impacts, it would be necessary to conduct tests with an artificial star with different simulations of the image quality. This is feasible with SPIRou, but requires more time.

\end{appendix}

\end{document}